\documentclass{jfm}
\usepackage[utf8]{inputenc}

\renewcommand{\pagelimitfooter}{}
\renewcommand{\absfooterflag}{}

\newcounter{ncorresp}
\let\Oldcorresp\corresp
\renewcommand{\corresp}[1]{{\Oldcorresp{#1}}\stepcounter{ncorresp}}

\usepackage{graphicx}
\usepackage{newtxtext}
\usepackage{newtxmath}
\usepackage{natbib}
\usepackage{hyperref}
\hypersetup{
    colorlinks = true,
    urlcolor   = blue,
    citecolor  = black,
}

\newcommand{\RomanNumeralCaps}[1]
\linenumbers

\usepackage{amsmath}
\usepackage{amssymb}
\usepackage{mathtools}
\usepackage{tikz}
\usepackage{aas_macros} 

\newcommand{\textcite}{\citet}
\newcommand{\Textcite}{\Citet}
\newcommand{\parencite}{\citep}

\renewcommand{\dh}{\partial}
\renewcommand{\d}{\mathrm{d}}

\newcommand{\meanBr}[1]{\left<#1\right>}
\newcommand{\dive}{\nabla\!\cdot}
\newcommand{\curl}{\nabla\!\times\!}

\newcommand{\Lap}{\nabla^2}

\newcommand{\defn}{\equiv}
\newcommand{\abs}[1]{\left|#1\right|}
\newcommand{\op}[1]{\operatorname{#1}}
\newcommand{\bigO}{O}

\newcommand{\scalesAs}{\sim}
\newcommand{\orderOf}{\sim}

\renewcommand{\vec}[1]{\boldsymbol{#1}}
\newcommand{\Heaviside}{\operatorname{\text{$\Theta$}}} 
\newcommand{\cross}{\times}
\newcommand{\Dirac}{\operatorname{\delta}}
\newcommand{\Rm}{\mbox{\textit{Rm}}} 
\newcommand{\St}{\mbox{\textit{St}}} 
\newcommand{\Pe}{\Pen} 

\newcommand{\continuedTerm}{\phantom{===}}
\newcommand{\negphantom}[1]{\settowidth{\dimen0}{#1}\hspace*{-\dimen0}} 
\newcommand{\emf}{\mathcal{E}}

\newcommand{\varied}[1]{\bar{#1}}
\newcommand*\circled[1]{\tikz[baseline=(char.base)]{\node[shape=circle,draw,inner sep=1pt] (char) {#1};}}
\newcommand{\posTimeComb}[1]{\circled{\scalebox{0.7}{\ensuremath{#1}}}}
\newcommand{\posTimeCombFree}{\maltese}
\newcommand{\ddxDummy}[1]{\eth^{(#1)}}
\newcommand{\velCorr}[1]{\left.\boxed{#1}\right.} 
\newcommand{\fnDer}[2]{\left.\boxed{\boxed{\frac{#1}{#2}}}\right.} 
\newcommand{\Sym}[1]{S_{#1}}
\newcommand{\derG}[2]{\Game_{#1}(#2)}

\newcommand{\funnyInt}{\oint}
\newcommand{\tcf}{\mathfrak{D}} 

\newcommand{\BbarVec}{\overline{\vec{B}}}
\newcommand{\Bbar}{\overline{B}}

\newcommand{\FT}[1]{\widetilde{#1}} 

\title{Turbulent transport in a non-Markovian velocity field}
\author{
G. Kishore
\corresp{\email{kishoreg@iucaa.in}}
\and
Nishant K. Singh
\corresp{\email{nishant@iucaa.in}}
}
\affiliation{Inter-University Centre for Astronomy \& Astrophysics, Post Bag 4, Ganeshkhind, Pune 411 007, India}

\begin{document}

\maketitle
\addtocounter{footnote}{\value{ncorresp}}

\begin{abstract}
	The commonly used quasilinear approximation allows one to calculate the turbulent transport coefficients for the mean of a passive scalar or a magnetic field in a given velocity field.
	Formally, the quasilinear approximation is exact when the correlation time of the velocity field is zero.
	We calculate the lowest-order corrections to the transport coefficients due to the correlation time being nonzero.
	For this, we use the Furutsu-Novikov theorem, which allows one to express the turbulent transport coefficients in a Gaussian random velocity field as a series in the correlation time.
	We find that the turbulent diffusivities of both the mean passive scalar and the mean magnetic field are suppressed.
	Nevertheless, contradicting a previous
	study,
	we show that the turbulent diffusivity of the mean magnetic field is smaller than that of the mean passive scalar.
	We also find corrections to the $\alpha$ effect.
\end{abstract}

\section{Introduction}

Astrophysical magnetic fields are observed on galactic, stellar, and planetary scales (\citealp[section 2]{KanduPhysicsReports2005}; \citealp{jones11}).
Stars such as the Sun exhibit periodic magnetic cycles, while
the Earth itself has a dipolar magnetic field that shields it from the solar wind.
Dynamo theory attempts to explain the generation and sustenance of such magnetic fields \citep{MoffattMagFieldGenBook, KrauseRadler80, KanduPhysicsReports2005, ShukurovKanduBook}.
Magnetic fields are often correlated at length scales much larger than that of the turbulent velocity field.
Mean-field magnetohydrodynamics takes advantage of this scale-separation to make the problem analytically tractable.

In general, the Lorentz force turns the evolution of the magnetic field into a nonlinear problem, which is difficult to study analytically.
As a first step, one can study the \emph{kinematic limit}, where the magnetic field is assumed to be so weak that the effect of the Lorentz force on the velocity field can be neglected.
The statistical properties of the velocity field can then be treated as given quantities, the effects of which on the magnetic field are to be determined.
In this study, we restrict ourselves to the kinematic dynamo.

Even in the kinematic limit, the evolution equation for the mean magnetic field depends on the correlation between the fluctuating velocity field and the fluctuating magnetic field, with the evolution equation for this correlation in turn depending on higher-order correlations (schematically, $\dh \meanBr{v^n B}/\dh t = \meanBr{v^{n+1} B}$ where $v$ is the velocity field and $B$ is the magnetic field).
To keep the system of equations manageable, one has to truncate this hierarchy by applying a \emph{closure}.
The most common closure in mean-field dynamo theory is the quasilinear approximation (also known as the First Order Smoothing Approximation, FOSA; or the Second Order Correlation Approximation, SOCA) (e.g.\@ \citealp[sec.~7.5]{MoffattMagFieldGenBook}; \citealp[sec.~4.3]{KrauseRadler80}), in which the evolution equation for the fluctuating magnetic field is linearized.
Strictly, this closure is valid at either low Reynolds number ($\Rey$, the ratio of the viscous timescale to the advective timescale) or low Strouhal number ($\St$, the ratio of the correlation time of the velocity field to its turnover time\footnote{
Note that this definition, which seems to be prevalent in the dynamo community \citep[going back to][eq.~3.14]{KrauseRadler80}, is different from the more common definition which is used for oscillatory flows \citep[e.g.][p.~295]{Whi99}.
}).
The former limit is astrophysically irrelevant.
The applicability of the latter limit can be judged from the fact that in simulations \citep{BraSub05, kapyla2006strouhal}, one typically finds $0.1 \leq \St \leq 1$.
While this suggests that the effects of a nonzero correlation time are not negligible, it leaves room for hope that perturbative approaches can at least capture the qualitative effects of having a nonzero correlation time.

A two-scale averaging procedure, where one performs successive averages over different spatiotemporal scales, is sometimes thought of as a simple device to go beyond the quasilinear approximation by capturing the effects of higher order correlations of the velocity field (e.g.\@ \citealp{kraichnan1976diffusion}; \citealp[p.~341]{silantev2000}).
This has been applied to passive-scalar and magnetic-field transport (\citealp{kraichnan1976diffusion}; \citealp[sec.~7.11]{MoffattMagFieldGenBook}; \citealp{singh2016moffatt}; \citealp{GopSin23}).
In particular, \textcite{kraichnan1976diffusion} has found that the turbulent diffusion of the mean passive scalar is not affected, while that of the mean magnetic field is suppressed.

More rigorous perturbative calculations have been performed by \textcite{knobloch77}, \textcite{drummond1982}, and \textcite{nicklaus88}.
Using independent approaches, \textcite[using the cumulant expansion]{knobloch77} and \textcite[using a path integral formalism]{drummond1982} have found that to the lowest order, the turbulent diffusion of the mean passive scalar is suppressed when the correlation time is nonzero.
Applying the cumulant expansion, \textcite{nicklaus88} have found that the turbulent diffusion of the magnetic field is also suppressed when the correlation time is nonzero.\footnote{
While \textcite{knobloch77} also treated the case of the mean magnetic field, that particular result has a problem which we point out in section \ref{section: comparison my mag results with cumulant expansion}.
}
As we will later see, comparison of the results obtained by \textcite{knobloch77} and \textcite{drummond1982} for the passive scalar with that reported by \textcite{nicklaus88} for the magnetic field suggests that the turbulent diffusivity for the mean magnetic field is \emph{identical} to that for the mean passive scalar even when the correlation time of the velocity field is nonzero.
This disagrees qualitatively with the findings of \textcite{kraichnan1976diffusion}.

\Citet{Miz23}, using a renormalization group analysis, has calculated the effect of the kinetic helicity on the diffusion of the mean magnetic field in the limit of low fractional helicity.
In their method, the correlation time of the velocity field is nonzero, and is implicitly determined by the equations of motion.
In qualitative agreement with the other studies mentioned above, they report that turbulent diffusion of the magnetic field is suppressed in a helical velocity field.
However, they have not considered the case of a passive scalar, and so it is still unclear if turbulent diffusion affects the mean passive scalar and the mean magnetic field in the same way.

For a Gaussian random velocity field, one can use the Furutsu-Novikov theorem \parencite{Fur63, novikov1965} to write turbulent transport coefficients as series in the correlation time of the velocity field.\footnote{
\Citet{SchekochihinKulsrud2001} discuss how this method is related to other methods such as the cumulant expansion.
}
While this approach has been used to study the small-scale dynamo \parencite{SchekochihinKulsrud2001, GopSin24}, passive scalar transport \parencite{gleeson2000closure}, and the effects of shear on plasma turbulence \parencite{ZhangMahajan2017}, we are not aware of it being used to study the transport of the mean magnetic field.

In this work, we use the Furutsu-Novikov theorem to calculate the lowest-order corrections (linear in the correlation time of the velocity field) to the transport coefficients for a mean passive scalar and a mean magnetic field.
For the mean passive scalar, we find that the turbulent diffusivity is suppressed, in agreement with previous work \parencite{drummond1982, knobloch77}.
We also find that turbulent diffusion of the mean magnetic field is suppressed more strongly than in the case of the mean passive scalar, disagreeing with the result obtained by \textcite{nicklaus88}.

In section \ref{passive-tensor: section: FOSA passive scalar}, we use the quasilinear approximation to study the diffusion of the mean passive scalar.
In section \ref{PS: section: FuruNovi corrections to FOSA}, we apply the Furutsu-Novikov theorem to the same problem, highlighting the differences as compared to the quasilinear approximation.
In section \ref{B: section: FuruNovi mean magnetic}, we apply the same technique to the turbulent transport of the mean magnetic field.
Finally, we summarize our conclusions in section \ref{passive-tensor: section: conclusions}.

\section{Scalar transport in the quasilinear approximation}
\label{passive-tensor: section: FOSA passive scalar}
\subsection{Mean and fluctuating fields}
We consider a passive scalar, evolving according to
\begin{equation}
	\frac{\dh\theta}{\dh t} = \kappa\nabla^2\theta - \vec{u}\cdot\nabla\theta
	\,.
	\label{PS: eq: theta evolution}
\end{equation}
We split the scalar into mean and fluctuating components, $\theta = \Phi + \phi$, choosing an averaging procedure, $\meanBr{\Box}$, that obeys Reynolds' rules \citep[e.g.][sec.~3.1]{MoninYaglomVol1}.
For simplicity, we assume $\meanBr{\vec{u}} = \vec{0}$.
The evolution equations for $\Phi$ and $\phi$ are
\begin{align}
	\frac{\dh\Phi}{\dh t}
	={}&
	\kappa\nabla^2\Phi
	- \meanBr{ \vec{u}\cdot\nabla\phi }
	\label{PS: eq: Phi evolution}
	\\
	\frac{\dh\phi}{\dh t}
	={}&
	\kappa\nabla^2\phi
	- \vec{u}\cdot\nabla\phi
	+ \meanBr{ \vec{u}\cdot\nabla\phi }
	- \vec{u}\cdot\nabla\Phi
	\,.
	\label{PS: eq: phi evolution}
\end{align}

\subsection{The quasilinear approximation}
In the quasilinear approximation, we discard second-order correlations of the fluctuating quantities in equation \ref{PS: eq: phi evolution}, and write
\begin{equation}
	\frac{\dh\phi}{\dh t}
	=
	\kappa\nabla^2\phi
	- \vec{u}\cdot\nabla\Phi
	\,.
\end{equation}
The above is an diffusion equation with a source term $-\vec{u}\cdot\nabla\Phi$.
Assuming the perturbations $\phi$ were zero at $t\to-\infty$,\footnote{
Note that under the quasilinear approximation, the initial condition does not contribute to $\meanBr{ \vec{u}\cdot\nabla\phi }$ at later times if it is uncorrelated with the fluctuating velocity field.
} we can write
\begin{equation}
	\phi(\vec{x},t) = - \int_{-\infty}^t\d\tau \int\d \vec{q} \, G(\vec{x},t|\vec{q}, \tau) \, \vec{u}\cdot\nabla_{\vec{q}}\Phi(\vec{q},\tau)
	\label{PS: eq: phi FOSA integral}
\end{equation}
where the diffusive Green function is
\begin{equation}
	G(\vec{x},t | \vec{x}', t')
	\defn
	\left[ 4\pi\kappa \left( t - t'\right)\right]^{-3/2} \exp{\!\left( - \frac{ \left( \vec{x} - \vec{x}'\right)^2 }{ 4\kappa \left( t - t' \right) } \right)}
	\,.
\end{equation}

We now consider the term $\meanBr{ \vec{u}\cdot\nabla\phi }$ in equation \ref{PS: eq: Phi evolution}.
Using equation \ref{PS: eq: phi FOSA integral}, we write this term as
\begin{equation}
	\meanBr{ \vec{u}\cdot\nabla\phi } = - \meanBr{ \vec{u}(\vec{x},t) \cdot \nabla_{\vec{x}} \int_{-\infty}^t\d\tau \int\d \vec{q} \, G(\vec{x},t|\vec{q}, \tau) \, \vec{u}(\vec{q},\tau)\cdot\nabla_{\vec{q}}\Phi(\vec{q},\tau) }
	\,.
	\label{PS: eq: turbulent diffusion term FOSA arbitrary velocity correlation compressible velocity}
\end{equation}
Assuming $\dive\vec{u}=0$, we write the above as
\begin{equation}
	\meanBr{ \vec{u}\cdot\nabla\phi } = - \frac{\dh}{\dh x_i} \int_{-\infty}^t\d\tau \int\d \vec{q} \, G(\vec{x},t|\vec{q}, \tau) \meanBr{ u_i(\vec{x},t) u_j(\vec{q},\tau) } \frac{\dh \Phi(\vec{q},\tau)}{\dh q_j}
	\,.
	\label{PS: eq: turbulent diffusion term FOSA arbitrary velocity correlation}
\end{equation}

\subsection{White-noise velocity field}
Let us assume a homogeneous, isotropic, delta-correlated velocity field, i.e.\@
\begin{equation}
	\meanBr{ u_i(\vec{x},t) u_j(\vec{x},\tau) } = \frac{2}{3} \, E  \, \delta_{ij} \,\delta(t-\tau)
	\,.
	\label{PS: eq: velocity field homogeneous delta correlated in space and time}
\end{equation}
Setting $\kappa = 0$ (so that we have $G(\vec{x},t|\vec{q}, \tau) = \delta(\vec{x}-\vec{q}) \Heaviside(t-\tau)$), equation \ref{PS: eq: turbulent diffusion term FOSA arbitrary velocity correlation} becomes
\begin{align}
	\begin{split}
		\meanBr{ \vec{u}\cdot\nabla\phi }
		={}&
		- \frac{\dh}{\dh x_i} \int_{-\infty}^t\d\tau \meanBr{ u_i(\vec{x},t) u_j(\vec{x},\tau) } \frac{\dh \Phi(\vec{q},\tau)}{\dh x_j}
	\end{split}
	\\
	\begin{split}
		={}&
		- \frac{E}{3} \, \nabla^2 \Phi
	\end{split}
\end{align}
where we have used $\int_0^\infty \Dirac(t) \, \d t = 1/2$.
Plugging the above into equation \ref{PS: eq: Phi evolution}, we obtain
\begin{equation}
	\frac{\dh\Phi}{\dh t}
	=
	\bigg( \kappa + \frac{E}{3} \bigg) \nabla^2\Phi
	\label{PS.FOSA: eq: Phi evolution final white noise}
\end{equation}
where $E/3$ is referred to as the turbulent diffusivity.
In this approximation, the only effect of the fluctuating velocity field is to enhance the diffusivity of the mean scalar field.

\subsection{Nonzero correlation time}
\label{PS.FOSA: section: nonzero correlation time}
Recall that equation \ref{PS: eq: turbulent diffusion term FOSA arbitrary velocity correlation} contains an integral over past values of $\Phi$.
If the correlation time of the velocity field (say $\tau_c$) is small, this integral can be converted to a series in $\tau_c$ by Taylor-expanding $\Phi(\vec{q}, \tau)$ about the time $t$.
Explicitly, expanding
\begin{equation}
	\Phi(\vec{q}, \tau)
	=
	\Phi(\vec{q}, t)
	+ \left( \tau - t \right) \frac{\dh \Phi(\vec{q}, t) }{\dh t}
	+ \bigO( (\tau - t)^2 )
\end{equation}
one writes\footnote{
In principle, one should also Taylor-expand the diffusive Green function which appears inside the integral, but here, we are only interested in the contributions that are independent of $\kappa$.
}
\begin{align}
	\begin{split}
		\meanBr{ \vec{u}\cdot\nabla\phi }
		={}&
		- \frac{\dh}{\dh x_i} \int_{-\infty}^t\d\tau \int\d \vec{q} \, G(\vec{x},t|\vec{q}, \tau) \meanBr{ u_i(\vec{x},t) u_j(\vec{q},\tau) } \frac{\dh \Phi(\vec{q},t)}{\dh q_j}
		\\& - \frac{\dh}{\dh x_i} \int_{-\infty}^t\d\tau \int\d \vec{q} \, G(\vec{x},t|\vec{q}, \tau) \meanBr{ u_i(\vec{x},t) u_j(\vec{q},\tau) } \left( \tau - t \right) \frac{\dh^2 \Phi(\vec{q},t)}{\dh q_j \dh t}
		\\& + \bigO{\!\left( \int_{-\infty}^t \d\tau \left( \tau - t\right)^2 \meanBr{u(\cdot,t) u(\cdot,\tau)} \right)}
		\,.
	\end{split} \label{PS.FOSA: taylor expansion Phi}
\end{align}
The three lines on the RHS are $\bigO(\tau_c^0)$, $\bigO(\tau_c^1)$, and $\bigO(\tau_c^2)$ respectively.
If one wishes to discard $\bigO(\tau_c^2)$ terms above, one can use equation \ref{PS.FOSA: eq: Phi evolution final white noise} to substitute for $\dh\Phi/\dh t$ appearing on the second line.

However, we shall soon see that the higher-order velocity correlations neglected in the quasilinear approximation also have $\bigO(\tau_c)$ contributions to the equation for $\meanBr{ \vec{u}\cdot\nabla\phi }$; one thus has to go beyond the quasilinear approximation to study the effects of having a nonzero correlation time.

\section{Scalar transport with nonzero correlation time}
\label{PS: section: FuruNovi corrections to FOSA}

\subsection{Application of the Furutsu-Novikov theorem}

We use the Furutsu-Novikov theorem (\citealp[eq.~5.18]{Fur63}; \citealp[eq.~2.1]{novikov1965}):
\begin{equation}
	\meanBr{ u_i(x,t) \lambda_i(x,t) } = \int \meanBr{u_i(x,t) u_j(x',t')} \meanBr{\frac{\delta \lambda_i(x,t)}{\delta u_j(x',t')}} \d^3 x'\d t' 
	\label{PS: eq: use FN theorem to write uGradphi term}
\end{equation}
where $u_i$ is a Gaussian random field with zero mean, and $\lambda_i$ is some functional of $u_i$.

We treat the evolution equation for the total passive scalar (equation \ref{PS: eq: theta evolution}) as diffusion with a source term $- \vec{u}\cdot\nabla\theta$ and write\footnote{
We have ignored a term containing the convolution of the Green function with the initial condition, since we expect its functional derivative wrt.\@ $\vec{u}$ at later times to be zero.
}
\begin{align}
	\theta(\vec{x},t) ={}& - \int_{-\infty}^t\d\tau \int\d \vec{q} \, G(\vec{x},t|\vec{q}, \tau) \, u_m(\vec{q},\tau)\frac{\dh\theta(\vec{q},\tau)}{\dh q_m}
	\label{PS: eq: theta as integral over source}
	\\\implies \frac{\dh\theta(\vec{x},t)}{\dh x_i} ={}& - \frac{\dh}{\dh x_i} \int_{-\infty}^t\d\tau \int\d \vec{q} \, G(\vec{x},t|\vec{q}, \tau) \, u_m(\vec{q},\tau)\frac{\dh\theta(\vec{q},\tau)}{\dh q_m} \label{PS: eq: implicit integral equation for lambda}
	\,.
\end{align}
Note that $G$ does not depend on $\vec{u}$.
Defining $\lambda_i \defn \dh_i\theta$ and taking the functional variation on both sides, we write
\begin{align}
	\delta\lambda_i(\vec{x},t) ={}& - \frac{\dh}{\dh x_i} \int_{-\infty}^t\d\tau \int\d \vec{q} \, G(\vec{x},t|\vec{q}, \tau) \left[ \lambda_m(\vec{q},\tau)\,\delta u_m(\vec{q},\tau) + u_m(\vec{q},\tau) \,\delta\lambda_m(\vec{q},\tau) \right]
	\\\begin{split}
		={}& - \frac{\dh}{\dh x_i} \int_{-\infty}^t\d\tau \int\d \vec{q} \, G(\vec{x},t|\vec{q}, \tau) \lambda_m(\vec{q},\tau)\,\delta u_m(\vec{q},\tau) 
		\\& - \frac{\dh}{\dh x_i} \int_{-\infty}^t\d\tau \int\d \vec{q} \d\vec{x}''\d t'' \, G(\vec{x},t|\vec{q}, \tau) \, u_m(\vec{q},\tau) \, \frac{\delta\lambda_m(\vec{q},\tau)}{\delta u_k(\vec{x''},t'')} \, \delta u_k(\vec{x''},t'') 
		\,.
	\end{split}
\end{align}
Using the above, we write
\begin{align}
	\begin{split}
		\frac{\delta\lambda_i(\vec{x},t)}{\delta u_j(\vec{x'},t')} ={}& - \frac{\dh}{\dh x_i} G(\vec{x},t|\vec{x}', t') \Heaviside(t-t') \, \lambda_j(\vec{x}',t')
		\\& - \frac{\dh}{\dh x_i} \int_{t'}^t\d\tau \int\d \vec{q} \, G(\vec{x},t|\vec{q}, \tau) \, u_m(\vec{q},\tau)  \,\frac{\delta\lambda_m(\vec{q},\tau)}{\delta u_j(\vec{x'},t')}  \Heaviside(t-t')
		\,.
	\end{split}
\end{align}
Averaging both sides,we write
\begin{align}
	\begin{split}
		\meanBr{ \frac{\delta\lambda_i(\vec{x},t)}{\delta u_j(\vec{x'},t')} } ={}& - \frac{\dh}{\dh x_i} G(\vec{x},t|\vec{x}', t') \Heaviside(t-t') \meanBr{ \lambda_j(\vec{x}',t') }
		\\& - \frac{\dh}{\dh x_i} \int_{t'}^t\d\tau \int\d \vec{q} \, G(\vec{x},t|\vec{q}, \tau) \Heaviside(t-t') \meanBr{ u_m(\vec{q},\tau) \,\frac{\delta\lambda_m(\vec{q},\tau)}{\delta u_j(\vec{x'},t')} }
		\,.
	\end{split} \label{PS: eq: delta lambda by delta uj with advection-diffusion Green function}
\end{align}
Now, we plug the above into equation \ref{PS: eq: use FN theorem to write uGradphi term} and write
\begin{align}
	\begin{split}
		& \meanBr{ \vec{u}\cdot\nabla\theta} = \int \meanBr{u_i(\vec{x},t) u_j(\vec{x}',t')} \meanBr{\frac{\delta \lambda_i(\vec{x},t)}{\delta u_j(\vec{x}',t')}} \d \vec{x}'\d t' 
	\end{split}
	\\\begin{split}
		&= - \int \d \vec{x}'\d t'  \meanBr{u_i(\vec{x},t) u_j(\vec{x}',t')} \frac{\dh}{\dh x_i} G(\vec{x},t|\vec{x}', t') \Heaviside(t-t') \meanBr{ \lambda_j(\vec{x}',t') }
		\\&\phantom{={}} - \int\d \vec{q}\, \d \vec{x}'\d t'  \int_{t'}^t\d\tau \meanBr{u_i(\vec{x},t) u_j(\vec{x}',t')} \Heaviside(t-t') \frac{\dh}{\dh x_i}  G(\vec{x},t|\vec{q}, \tau) \meanBr{ u_m(\vec{q},\tau)  \,\frac{\delta\lambda_m(\vec{q},\tau)}{\delta u_j(\vec{x'},t')} }
		\,.
	\end{split}
\end{align}
The first term in the above is exactly the expression obtained using the quasilinear approximation (equation \ref{PS: eq: turbulent diffusion term FOSA arbitrary velocity correlation compressible velocity}).
The second term becomes zero if the correlation time of the velocity field is zero, and can be thought of as a correction to the quasilinear approximation.

In fact, by repeated application of the Furutsu-Novikov theorem, the second term can be expanded as a series.
To easily represent the terms of this series, we now define some new symbols.
We use $\posTimeComb{a}$, where $a=1,2,\dots$, to denote a combination of position and time variables which are integrated over, so that, e.g., $\vec{u}(\vec{x}^{(1)}, t^{(1)})$ stands for $\vec{u}\posTimeComb{1}$.
A derivative wrt.\@ the position variable labelled by $a$ is denoted by $\ddxDummy{a}$.
$\posTimeCombFree$ denotes the special combination $(\vec{x},t)$, which is not integrated over.
We use $\velCorr{1_i2_j}$ to denote $\meanBr{u_i\posTimeComb{1} u_j\posTimeComb{2}}$.
Further,
\begin{align}
	\begin{split}
		\fnDer{\beta_m}{1_{j_1}\dots n_{j_n}}
		\defn{}&
		\meanBr{ \frac{\delta^{n} \lambda_m\posTimeComb{\beta}}{\delta u_{j_1}\posTimeComb{1} \dots u_{j_n} \posTimeComb{n} } }
	\end{split}
	\\
	\begin{split}
		\derG{m}{1|2}
		\defn{}&
		\ddxDummy{1}_m G^c\!(\posTimeComb{1}|\posTimeComb{2})
	\end{split}
	\label{PS: eq: funny notation definition}
\end{align}
where $G^c$ is the causal Green function (defined such that $G^{c}\!(x,t|x',t') = 0$ if $t'>t$; $G(x,t|x',t')$ otherwise).
In this notation, the $n$-th functional derivative of $\lambda$ (the expression for which is derived in appendix \ref{PS: appendix: nth functional derivative recursion-like relation} as equation \ref{PS: eq: nth functional derivative recursion-like relation}) can be written as
\begin{align}
	\begin{split}
		\fnDer{\posTimeCombFree_i}{ 1_{j_1} \dots n_{j_n} }
		={}& - \Sym{1,2,\dots,n} \derG{i}{\posTimeCombFree|1} \fnDer{ 1_{j_1} }{ 2_{j_2}\dots n_{j_n} }
		\\& - \int\d A\d B\, \derG{i}{\posTimeCombFree|A} \velCorr{ A_m B_a } \fnDer{A_m}{B_a 1_{j_1}\dots n_{j_n} }
	\end{split} \label{PS: eq: nth functional derivative recursion-like relation in funny notation}
\end{align}
where $\Sym{abc\dots}$ is an operator that symmetrizes over all the arguments indicated in the subscript.\footnote{
E.g., $S_{12} A_{12} = A_{12} + A_{21}$.
}
Hereafter, we will dispense with integral symbols, proceeding with the understanding that $\posTimeCombFree$ is the only variable that is not integrated over.
Applying equation \ref{PS: eq: nth functional derivative recursion-like relation in funny notation} thrice to equation \ref{PS: eq: use FN theorem to write uGradphi term} and discarding terms with more than two velocity correlations, we are left with
\begin{align}
	\begin{split}
		\meanBr{ \vec{u}\cdot\nabla\theta} 
		={}&
		\velCorr{\posTimeCombFree_i 1_{i_1} } \fnDer{ \posTimeCombFree_i}{1_{i_1}}
		\label{PS: eq: FuruNovi series object before applying recursion relation}
	\end{split}
	\\
	\begin{split}
		={}&
		- \velCorr{{\posTimeCombFree}_{i} {1}_{i_{1}}} \derG{i}{\posTimeCombFree|1} \meanBr{\lambda_{i_{1}}\posTimeComb{1} }
		\\& - \velCorr{{\posTimeCombFree}_{i} {1}_{i_{1}}} \derG{i}{\posTimeCombFree|2} \velCorr{{2}_{i_{2}} {3}_{i_{3}}} \derG{i_{2}}{2|1} \derG{i_{1}}{1|3} \meanBr{\lambda_{i_{3}}\posTimeComb{3} }
		\\& - \velCorr{{\posTimeCombFree}_{i} {1}_{i_{1}}} \derG{i}{\posTimeCombFree|2} \velCorr{{2}_{i_{2}} {3}_{i_{3}}} \derG{i_{2}}{2|3} \derG{i_{3}}{3|1} \meanBr{\lambda_{i_{1}}\posTimeComb{1} }
		\,.
	\end{split} \label{PS: eq: series with terms up to v4}
\end{align}
As pointed out earlier, the first term above is just the quasilinear term (see equation \ref{PS: eq: turbulent diffusion term FOSA arbitrary velocity correlation compressible velocity}).
To consistently obtain the $\bigO(\tau_c)$ contributions to the above equation, one should also expand the quasilinear term as a series in the correlation time, as described in section \ref{PS.FOSA: section: nonzero correlation time}.

\subsection{The order of the neglected terms}

The leading-order (in $\tau_c$) contribution of a particular term can be deduced as follows.\footnote{
A term whose leading-order contribution is at a particular order in $\tau_c$ may contain higher-order contributions as well, as discussed in section \ref{PS.FOSA: section: nonzero correlation time}.
}
Each velocity correlation introduces a factor of $\tcf$ (see equation \ref{PS: eq: separable correlator definition}), for which $\int \tcf(\tau)\d\tau = \bigO(1)$.
Each of the remaining time integrals contributes a factor of $\tau_c$.
The order of a particular term in $\tau_c$ is then simply the number of time integrals minus the number of velocity correlations.
The object on the RHS of equation \ref{PS: eq: FuruNovi series object before applying recursion relation} contains one time integral with one velocity correlation, so it has no `explicit' factors of $\tau_c$ (everything is hidden inside the functional derivative).

\begin{figure}
	\centering
	\includegraphics[width=0.9\textwidth,keepaspectratio=true]{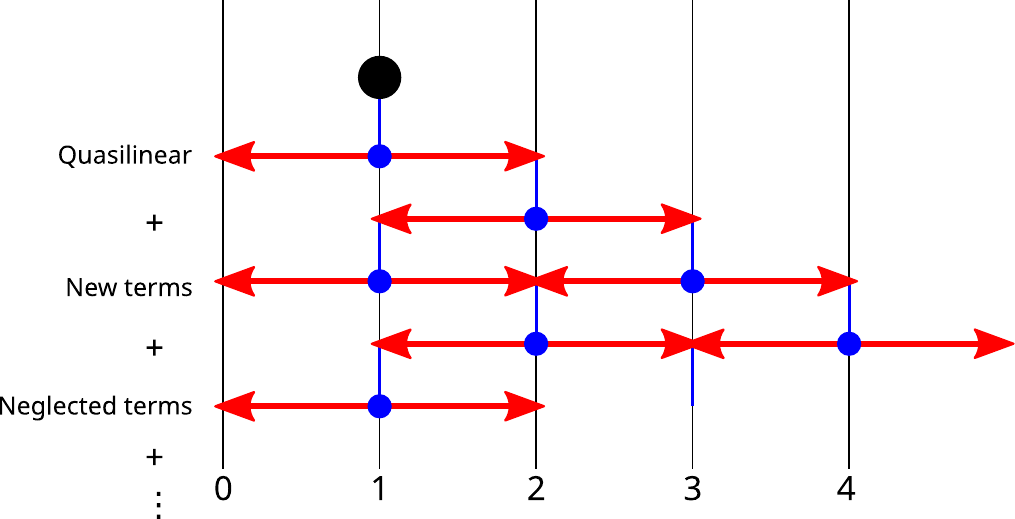}
	\caption{
	Diagram used to find the order in $\tau_c$ of the neglected terms.
	}
	\label{PS: fig: FuruNovi series pictorial explanation}
\end{figure}
In figure \ref{PS: fig: FuruNovi series pictorial explanation}, the vertical lines denote functional derivatives of various orders. 
Starting from the RHS of equation \ref{PS: eq: FuruNovi series object before applying recursion relation} (black circle), one obtains various terms after repeated application of the recursion relation (equation \ref{PS: eq: nth functional derivative recursion-like relation} or \ref{PS: eq: nth functional derivative recursion-like relation in funny notation}; in the figure, each blue circle with arrows leading out corresponds to one application of the relation).
A rightward step adds two time integrals and a velocity correlation, while a leftward step introduces neither velocity correlations nor time integrals.
The order of a particular term in $\tau_c$ is thus simply the number of rightward steps taken to reach it, starting from the black circle.
The terms we have neglected are $\bigO(\tau_c^2)$.

\subsection{Simplification assuming separable, weakly inhomogeneous turbulence}

\subsubsection{Assumptions}
Assuming $\kappa=0$ (which corresponds to $\Pe \gg 1$; $\Pe$, the \emph{Peclet number}, is the ratio of the diffusive timescale to the advective timescale for the total passive scalar),
we simplify the terms on the RHS of equation \ref{PS: eq: series with terms up to v4} in appendix \ref{PS: appendix: simplify v4 terms alt method}.
The expressions obtained are somewhat long.
We proceed by assuming the velocity correlations are separable, i.e.\@ 
\begin{equation}
	\meanBr{u_i(\vec{x},t+\tau) \dh^n_{i_1\dots i_n} u_j(\vec{x},t)} = C_{iji_1\dots i_n}(\vec{x},t) \tcf(\tau)
	\,.
	\label{PS: eq: separable correlator definition}
\end{equation}
The temporal correlation function, $\tcf$, has the properties
\begin{equation}
	2 \int_0^\infty \tcf(t) \, \d t = 1
	\,,\quad
	2 \int_0^\infty t \, \tcf(t) \, \d t = \tau_c
	\label{PS: eq: tcf properties}%
	\,,
\end{equation}
where $\tau_c$ is the correlation time.
Note that we have assumed the correlation time is independent of the spatial scale.
We further assume that $C(\vec{x},t)$ and $\meanBr{\lambda_i(\vec{x},t)}$ vary on a timescale much larger than the timescale of $\tcf(\tau)$.
Discarding terms with more than two derivatives of the mean scalar field ($\Phi$) and taking the `weakly inhomogeneous' case (where we keep only one large-scale derivative of the turbulent spectra; see appendix \ref{PS: appendix: weakly inhomogeneous correlators}), we obtain simple expressions, which we describe below.

\subsubsection{The diffusion equation for the mean scalar}


%
%
%
%
%

\begin{table}
	\centering
	\begingroup
		\renewcommand{\arraystretch}{2}
		\begin{tabular}{llll}
			Name & $\tcf(\tau)$ &  $g_2$ \\
			\hline
			Exponential & $\displaystyle \frac{1}{2\tau_c} \, e^{-\abs{\tau}/\tau_c} $ & 1/8 \\
			Top hat & $\displaystyle \frac{1}{4\tau_c} \, \Heaviside{\!\left( 2\tau_c - \tau \right)} \Heaviside{\!\left( \tau + 2 \tau_c \right)}$ & 1/12
		\end{tabular}
	\endgroup
	\caption[
		Values of $g_2$ for some temporal correlation functions.
		]{
		Values of $g_2$ for some temporal correlation functions.
		The corresponding values of $g_1$ can be calculated using equation \ref{B2.SSD.real: eq: g1 + g2}.
		The constants in the correlation functions have been chosen to satisfy equations \ref{PS: eq: tcf properties}.
		}
	\label{PS: table: g2 values}
\end{table}

Setting $\kappa=0$, the average of the evolution equation for the total passive scalar (equation \ref{PS: eq: theta evolution})
\begin{equation}
	\frac{\dh\Phi}{\dh t} = - \meanBr{\vec{u}\cdot\nabla\theta}
\end{equation}
Using equations \ref{PS: eq: post-FOSA O(K) term 1}, \ref{PS: eq: post-FOSA O(K) term 2}, and \ref{PS: eq: FOSA O(K) term} (appendix \ref{PS: appendix: simplify separable correlator}), we write the above as
\begin{equation}
	\frac{\dh\Phi}{\dh t} = \frac{\dh}{\dh x_i}{\left[ \left( \frac{E}{3} + \frac{\tau_c g_1 H^2 }{18} - \frac{\tau_c 2 g_2 E N}{9} \right) \frac{\dh \Phi}{\dh x_i} \right]}
	\,.
	\label{PS.FN: eq: dPhidt final general tcf}
\end{equation}
where $g_1$ and $g_2$ depend on the temporal correlation function $\tcf(\tau)$ (table \ref{PS: table: g2 values} lists some examples), and
\begin{subequations}
	\begin{align}
		E(\vec{x},t) \defn{}& \int_0^\infty \meanBr{ u_i(\vec{x},t+\tau)u_i(\vec{x},t) } \d\tau
		\label{PS: eq: E defn}
		\\
		H(\vec{x},t) \defn{}& 2 \int_0^\infty \meanBr{ u_i(\vec{x},t+\tau)\omega_i(\vec{x},t) } \d\tau
		\\
		N(\vec{x},t) \defn{}& 2 \int_0^\infty \meanBr{ \omega_i(\vec{x},t+\tau)\omega_i(\vec{x},t) } \d\tau
		\,.
	\end{align} \label{PS: eq: E H N definitions}
\end{subequations}
Note that the definition of $E$ is consistent with that in equation \ref{PS: eq: velocity field homogeneous delta correlated in space and time}.

For a maximally helical velocity field with a top-hat temporal correlation function ($g_2 = 1/12$ and $H^2 = 2EN$), the $\bigO(\tau_c)$ correction to the turbulent diffusivity becomes zero.
On the other hand, if the temporal correlation function is exponential, we obtain
\begin{equation}
	\frac{\dh\Phi}{\dh t} = \frac{\dh}{\dh x_i}{\left[ \left( \frac{E}{3} - \frac{\tau_c }{144} \left( 4 E N - H^2 \right) \right) \frac{\dh \Phi}{\dh x_i} \right]}
	+ \bigO(\tau_c^2)
	\,.
	\label{PS: eq: negative diffusivity from FuruNovi3, weakly inhomogeneous, kappa and U equal 0}
\end{equation}
Note that the $\bigO(\tau_c)$ terms in this case cannot be zero, since the Cauchy-Schwarz inequality guarantees that $H^2 \le 2EN$.
The corrections above match those obtained by \textcite[eq.~5.1]{drummond1982} and \textcite[eq.~38]{knobloch77}.\footnote{
In both cases, we compared results after assuming an exponential temporal correlation function.
The expressions given by \textcite{knobloch77} need to be further simplified assuming a statistically homogeneous and isotropic Gaussian velocity field.
}

Examination of equation \ref{PS.FN: eq: dPhidt final general tcf} shows that for weakly inhomogeneous turbulence, the effective diffusivity becomes negative when $\tau_c N (2 g_2 - g_1 f^2) > 3$ (where we have defined $f^2 \defn H^2/2EN \in [0,1]$) and is positive otherwise.
However, when the $\bigO(\tau_c)$ corrections are so strong, one would expect the neglected $\bigO(\tau_c^2)$ terms to also become non-negligible, making equation \ref{PS.FN: eq: dPhidt final general tcf} invalid.

\subsubsection{Validity of the expansion}
\label{section: estimate tau*N}

We now make a crude estimate of how $\tau_c N$, which controls the validity of equation \ref{PS: eq: negative diffusivity from FuruNovi3, weakly inhomogeneous, kappa and U equal 0}, depends on $\Rey$.
Assuming the velocity field is homogeneous and isotropic, we estimate (recall that $N$ was defined in equation \ref{PS: eq: E H N definitions})
\begin{equation}
	N \propto \int k^2 \FT{C}_{ii}(k) \, \d \vec{k}
\end{equation}
where $\FT{C}_{ii}$ is the Fourier transform of $C_{ii}$ (equation \ref{PS: eq: separable correlator definition}).
Let us assume $\FT{C}_{ii}(k)$ is a power law in $k_0 < k < k_\nu$ (which we refer to as the inertial range) and zero elsewhere.
If we assume the velocity field obeys the Kolmogorov scaling relations \citep[e.g.][section 1.6]{davidson2004turbulence} in the inertial range, the fact that  $C_{ii}(\vec{x})$ is dimensionally a diffusivity implies $\FT{C}_{ii}(k) \scalesAs k^{-13/3}$.
We then estimate
\begin{equation}
	N
	\propto \int_{k_0}^{k_\nu} k^{-1/3} \d k
	\propto k_\nu^{2/3} - k_0^{2/3}
	\approx k_\nu^{2/3}
\end{equation}
where the last step assumes $k_\nu \gg k_0$.
Similarly, we can also estimate $E \scalesAs k_0^{-4/3}$, which gives us
$
	N/(E k_0^2)
	\propto
	\Rey^{1/2}
$.
Defining $\St \defn \tau_c E k_0^2$, we then find
\begin{equation}
	\tau_c N
	\propto
	\St \, \Rey^{1/2}
	\,.
\end{equation}
This means that for equation \ref{PS: eq: negative diffusivity from FuruNovi3, weakly inhomogeneous, kappa and U equal 0} to be valid, we require both the Strouhal number and the Reynolds number to not be large.
We believe the latter limitation is due to our assumption that the velocity field at all scales can be characterized by a single scale-independent correlation time ($\tau_c$).

\subsection{Aside: validity of the quasilinear approximation}

Let us now try to understand the regimes in which the quasilinear approximation is valid.
From the discussion above, it is clear that the quasilinear approximation becomes valid when the correlation time of the velocity field approaches zero.

There is also another limit in which the quasilinear approximation is valid.
Even when the correlation time is nonzero, one may expect the quasilinear approximation to remain valid if the $\bigO(\tau_c)$ corrections from the quasilinear approximation (section \ref{PS.FOSA: section: nonzero correlation time}) are much larger than the $\bigO(\tau_c)$ contributions from the second and third terms on the RHS of equation \ref{PS: eq: series with terms up to v4}.
We may estimate the ratios of these contributions as
\begin{equation}
	\frac{\text{post-quasilinear} }{\text{quasilinear} }
	\orderOf 
	\left. \left( \frac{\tau^3 u^4}{l^4} \Phi \right) \middle/ \left( \frac{\tau^2 u^2}{l^2} \, \frac{\dh \Phi}{\dh t} \right) \right.
\end{equation}
where $l$ denotes a length scale typical of the spatial derivative of an averaged quantity, $\tau$ denotes the correlation time of the fluctuating velocity field, and $u$ is the RMS value of the fluctuating velocity field.
If we further use equation \ref{PS.FOSA: eq: Phi evolution final white noise} for $\dh \Phi/\dh t$, we can write
\begin{equation}
	\frac{\text{post-quasilinear} }{\text{quasilinear} }
	\orderOf \frac{ \tau u^2 }{ \kappa  + \tau u^2/3}
	\,.
\end{equation}
Estimating $\tau \orderOf l/u$ (i.e.\@ $St \orderOf 1$), we find
\begin{equation}
	\frac{\text{post-quasilinear} }{\text{quasilinear} }
	\orderOf \frac{ \Pe }{ 1 + \Pe/3 }
	\,.
\end{equation}
where $\Pe \defn ul/\kappa$.
We thus see that even if the correlation time of the velocity field is not small, the quasilinear approximation can remain valid as long as $\Pe$ is small.

\section{Magnetic field transport with nonzero correlation time}
\label{B: section: FuruNovi mean magnetic}

\subsection{Application of the Furutsu-Novikov theorem}
Let us consider the induction equation with constant $\eta$:
\begin{equation}
	\frac{\dh\vec{B}}{\dh t} = \curl\left( \vec{u}\cross\vec{B} \right) + \eta \nabla^2 \vec{B}
	\,.
	\label{B.FN: eq: induction}
\end{equation}
For simplicity, we assume $\meanBr{\vec{u}} = \vec{0}$.
Averaging both sides, we find that the equation for the mean magnetic field is
\begin{equation}
	\frac{\dh \BbarVec}{\dh t} = \curl\vec{\emf} + \eta \Lap\BbarVec
	\label{B.FN: eq: mean field induction general EMF}
\end{equation}
where
\begin{equation}
	\vec{\emf} \defn \meanBr{ \vec{u} \cross \vec{b}}
	\,,\quad \vec{b} \defn \vec{B} - \BbarVec{}
	\,.
\end{equation}
To solve for the evolution of the mean magnetic field without solving for the fluctuating fields, we require an expression for $\vec{\emf}$ that depends only on known statistical properties of the velocity field and on $\BbarVec$.

Treating the first term on the RHS of equation \ref{B.FN: eq: induction} as a source, we can write the magnetic field at some arbitrary time as (analogous to equation \ref{PS: eq: theta as integral over source} for the passive scalar)
\begin{equation}
	B_i(\vec{x},t) = \int_{-\infty}^t\d\tau\int\d\vec{q}\, G(\vec{x},t|\vec{q},\tau) \, \epsilon_{ijk} \epsilon_{klm} \frac{\dh}{\dh q_j} \left( u_l(\vec{q},\tau) B_m(\vec{q},\tau) \right)
	\,.
	\label{B: eq: B_i general equation in terms of GF}
\end{equation}

We assume $\vec{u}$ is a Gaussian random field.
Recalling that $\meanBr{\vec{u}} = \vec{0}$, the EMF can be written as $\vec{\emf} = \meanBr{\vec{u} \cross \vec{B}}$.
The Furutsu-Novikov theorem (equation \ref{PS: eq: use FN theorem to write uGradphi term}) takes the form
\begin{align}
	\begin{split}
		\meanBr{u_i(\vec{x},t) B_j(\vec{x},t)} = \int \d \vec{x}'\d t' \meanBr{u_i(\vec{x},t) u_k(\vec{x}',t')} \meanBr{\frac{\delta B_j(\vec{x},t)}{\delta u_k(\vec{x}',t')}} 
		\,.
	\end{split} \label{B: eq: VB correlator starting expression before applying FuruNovi series}
\end{align}
For the sake of brevity, we will use notation similar to that described on page \pageref{PS: eq: funny notation definition}.
Explicitly,
\begin{align}
	\begin{split}
		\fnDer{\beta_m}{1_{j_1}\dots n_{j_n}}
		\defn{}&
		\meanBr{ \frac{\delta^{n} \lambda_m\posTimeComb{\beta}}{\delta u_{j_1}\posTimeComb{1} \dots u_{j_n} \posTimeComb{n} } }
	\end{split}
	\\
	\begin{split}
		\derG{m}{1|2}
		\defn{}&
		\ddxDummy{2}_m G^c\!(\posTimeComb{1}|\posTimeComb{2})
	\end{split}
	\\
	\begin{split}
		\Upsilon_{ijlm}
		\defn{}&
		\epsilon_{ijk}\epsilon_{klm}
		\,.
	\end{split}
\end{align}
The average of the $n$-th functional derivative of $B$ (derived in appendix \ref{B: appendix: nth functional derivative of B} as equation \ref{B: eq: n-th functional derivative of B recursion-like relation}) can then be written as
\begin{align}
	\begin{split}
		\fnDer{\posTimeCombFree_i}{1_{i_1}\dots n_{i_n}}
		={}& - \int\d A\d B \, \derG{j}{\posTimeCombFree|\posTimeComb{A}} \Upsilon_{ijlm} \velCorr{A_l B_a} \fnDer{A_m}{1_{i_1}\dots n_{i_n} B_a } 
		\\& - \sum_{\alpha=1}^n \derG{j}{\posTimeCombFree|\posTimeComb{\alpha}} \Upsilon_{ij i_{\alpha}m}  \fnDer{\alpha_m}{1_{i_1} \dots (\alpha-1)_{i_{\alpha-1}} (\alpha+1)_{i_{\alpha+1}} \dots n_{i_n} }
		\,.
	\end{split} \label{B: eq: n-th functional derivative recursion relation in funny notation}
\end{align}
Henceforth, we adopt the convention that $\posTimeCombFree$ is the only variable that is not integrated over.
Equation \ref{B: eq: VB correlator starting expression before applying FuruNovi series} can be written as
\begin{align}
	\begin{split}
		\continuedTerm&\negphantom{\continuedTerm}
		\meanBr{u_i(\vec{x},t) B_j(\vec{x},t)}
		=
		\velCorr{\posTimeCombFree_i 1_k} \fnDer{\posTimeCombFree_j}{1_k}
	\end{split}
	\\
	\begin{split}
		={}&
		- \Upsilon_{j i_{2} i_{1} i_{3}} \velCorr{{\posTimeCombFree}_{i} {1}_{i_{1}}} \derG{i_{2}}{\posTimeCombFree|1} \Bbar_{i_{3}}\posTimeComb{1}
		\\&
		- \Upsilon_{i_{5} i_{6} i_{1} i_{7}} \Upsilon_{j j_{3} i_{2} j_{2}} \Upsilon_{j_{2} i_{4} i_{3} i_{5}} \velCorr{{\posTimeCombFree}_{i} {1}_{i_{1}}} \derG{j_{3}}{\posTimeCombFree|2} \velCorr{{2}_{i_{2}} {3}_{i_{3}}} \derG{i_{4}}{2|3} \derG{i_{6}}{3|1} \Bbar_{i_{7}}\posTimeComb{1} 
		\\&
		- \Upsilon_{i_{5} i_{6} i_{3} i_{7}} \Upsilon_{j j_{3} i_{2} j_{2}} \Upsilon_{j_{2} i_{4} i_{1} i_{5}} \velCorr{{\posTimeCombFree}_{i} {1}_{i_{1}}} \derG{j_{3}}{\posTimeCombFree|2} \velCorr{{2}_{i_{2}} {3}_{i_{3}}} \derG{i_{4}}{2|1} \derG{i_{6}}{1|3} \Bbar_{i_{7}}\posTimeComb{3} 
	\end{split} \label{B: eq: vb correlator after applying FuruNovi thrice, mean velocity zero}
\end{align}
where we have applied the relation \ref{B: eq: n-th functional derivative recursion relation in funny notation} thrice and discarded terms with more than two velocity correlations.
As argued in the case of the passive scalar (section \ref{PS: section: FuruNovi corrections to FOSA}), the first term is $\bigO(\tau_c^0)$ (but also contains $\bigO(\tau_c)$ contributions), while the next two terms are $\bigO(\tau_c)$, where $\tau_c$ is the correlation time of the velocity field.

\subsection{The white-noise limit}
The $\bigO(\tau_c^0)$ term of equation \ref{B: eq: vb correlator after applying FuruNovi thrice, mean velocity zero} is
\begin{align}
	\meanBr{u_i B_j}
	={}& - \Upsilon_{j i_{2} i_{1} i_{3}} \velCorr{{\posTimeCombFree}_{i} {1}_{i_{1}}} \derG{i_{2}}{\posTimeCombFree|1} \Bbar_{i_{3}}\posTimeComb{1}
	\\\begin{split}
		={}& - \Upsilon_{j i_{2} i_{1} i_{3}} \int\d\vec{x}'\d t' \meanBr{ u_{i}(\vec{x},t) u_{i_{1}}(\vec{x}',t') } \frac{\dh G^c(\vec{x},t|\vec{x}',t') }{\dh x'_{i_2}} \Bbar_{i_{3}}(\vec{x}',t') 
		\,.
	\end{split}
\end{align}
Assuming $\eta=0$ and discarding the $\bigO(\tau_c)$ parts of this term, we find that
\begin{align}
	\begin{split}
		\meanBr{u_i B_j}
		={}& \Upsilon_{j i_{2} i_{1} i_{3}} \int_{-\infty}^t \d t'\, \meanBr{ u_{i}(\vec{x},t) \frac{\dh}{\dh x_{i_2}} u_{i_{1}}(\vec{x},t')  } \Bbar_{i_{3}}(\vec{x},t)
		\\& + \Upsilon_{j i_{2} i_{1} i_{3}} \int_{-\infty}^t\d t'\, \meanBr{ u_{i}(\vec{x},t) u_{i_{1}}(\vec{x},t') } \frac{\dh \Bbar_{i_{3}}(\vec{x},t) }{\dh x_{i_2}}
		\,.
	\end{split}
\end{align}
Assuming the velocity correlations are separable (equation \ref{PS: eq: separable correlator definition}) and the turbulence is homogeneous, we can use the expressions from appendix \ref{PS: appendix: unequal-time-correlators separable} and write
\begin{align}
	\begin{split}
		\meanBr{V_i B_j}
		={}&
		\frac{1}{12} \epsilon_{i i_2 j} H \Bbar_{i_2}
		- \frac{E}{3} \, \frac{\dh \Bbar_j }{\dh x_i}
	\end{split}
\end{align}
where we have used the fact that the divergence of the magnetic field is zero.
The EMF can then be written as
\begin{equation}
	\emf_k = \epsilon_{kij}\meanBr{V_i B_j}
	= - \frac{H}{6} \, \Bbar_{k}
	- \frac{E}{3} \, \epsilon_{kij} \dh_i \Bbar_j
	\label{B.FN: eq: white noise EMF}
\end{equation}
which is exactly the same as the usual quasilinear expression \parencite[e.g.][chapter 7]{MoffattMagFieldGenBook}.
This is usually written in the form
\begin{equation}
	\emf_k
	=
	\alpha \, \Bbar_{k}
	- \eta \, \epsilon_{kij} \dh_i \Bbar_j
	\,.
	\label{B.FN: eq: white noise EMF in terms of alpha eta}
\end{equation}
The coefficient $\alpha$ describes how a helical velocity field can drive the growth of a mean magnetic field, while the coefficient $\eta$ (called the turbulent diffusivity) describes dissipation of a mean magnetic field through the action of turbulence.

\subsection{Corrections due to nonzero correlation time}

\subsubsection{Expression for the EMF}

In appendix \ref{B: appendix: simplify FuruNovi O(tau) terms}, we simplify all the three terms on the RHS of equation \ref{B: eq: vb correlator after applying FuruNovi thrice, mean velocity zero}, keeping $\bigO(\tau_c)$ contributions, assuming the turbulence is homogeneous and isotropic, and setting $\eta=0$.
Setting $\eta=0$ in these terms corresponds to assuming $\Rm \gg 1$ ($\Rm$, the \emph{magnetic Reynolds number}, is the ratio of the diffusive timescale to the advective timescale for the total magnetic field).
The overall contribution to the EMF ($\emf_k = \epsilon_{kij} \left< V_i B_j \right>$, the contributions to which are given by equations \ref{B: eq: FOSA O(tau) term emf contribution simplified}, \ref{B: eq: Furunovi O(tau) term 1 simplified} and \ref{B: eq: Furunovi O(tau) term 2 simplified}) is
\begin{align}
	\begin{split}
		\emf_k
		={}&
		\left( \alpha_0 + \tau_c \alpha_1 \right) \Bbar{}_{k}
		- \left( \eta_0 - \tau_{c} \eta_1 \right) \epsilon{}_{kr_{0}r_{1}} \dh_{r_0} \Bbar{}_{r_{1}}
		\,.
		\label{B.FN: emf O(tau) final general tcf}
	\end{split}
\end{align}
Above, $\alpha_0$ and $\eta_0$ are the coefficients when the correlation time is zero (equation \ref{B.FN: eq: white noise EMF in terms of alpha eta});
\begin{align}
	\begin{split}
		\alpha_1
		\defn{}&
		\frac{ g_2 }{9} \left( 2 E L + H N \right)
		\,,
	\end{split}
	\\
	\begin{split}
		\eta_1
		\defn{}&
		\frac{2 E N g_{2}}{9} + \frac{H^{2} g_{2}}{18} + \frac{H^{2}}{24}
		\,;
	\end{split}
\end{align}
we have eliminated $g_1$ by using the fact that $g_1 + g_2 = 1/4$ (equation \ref{B2.SSD.real: eq: g1 + g2});
$E$, $H$, and $N$ are defined in equations \ref{PS: eq: E H N definitions}; and
\begin{equation}
	L(\vec{x},t) \defn 2 \int_0^\infty \meanBr{\vec{\omega}(\vec{x},t+\tau) \cdot \left[ \curl\vec{\omega}(\vec{x},t) \right] } \d\tau
	\,.
	\label{PS: eq: L definition}
\end{equation}
Note that the turbulent diffusivity is always reduced by the $\bigO(\tau_c)$ terms.
The validity of this equation is also determined by the criterion given in section \ref{section: estimate tau*N}.

Denoting the turbulent diffusivity for the mean magnetic field as $\eta_B$, and that for the mean passive scalar (equation \ref{PS.FN: eq: dPhidt final general tcf}) as $\eta_\theta$, we find that
\begin{equation}
	\eta_B - \eta_\theta
	=
	- \tau_c H^2 \left( \frac{g_1 + g_2}{18} + \frac{1}{24} \right)
	=
	- \frac{\tau_c H^2}{18}
\end{equation}
where we have used $g_1 + g_2 = 1/4$ (equation \ref{B2.SSD.real: eq: g1 + g2}).
Note that this is independent of the form of the temporal correlation function of the velocity field.

\subsubsection{Comparison with the cumulant expansion}
\label{section: comparison my mag results with cumulant expansion}

\Citet{knobloch77} and \citet{nicklaus88} have also studied such corrections using the cumulant expansion.\footnote{
\Textcite[p.~155]{nicklaus88} point out some issues with the calculation reported by \textcite{knobloch77}.
We agree with the misprint they have pointed out in Knobloch's equation A.13.
However, we agree with Knobloch's equation A.8.
We have not attempted to verify Knobloch's equation 42.
}\footnote{
\Textcite[sec.~II D]{SchekochihinKulsrud2001} have shown that at least for a simple model problem, the cumulant expansion is consistent with the method we have used.
}
In particular, \textcite[eqs.~24,26]{nicklaus88} have further simplified their results by treating the velocity field as a Gaussian random field.
Our results are expected to agree with theirs for $t \gg \tau_c$.
Our corrections to the $\alpha$ effect indeed match theirs.
However, their stated correction to the turbulent diffusivity of the magnetic field differs from our result above, and is instead identical to what we had obtained for passive scalar diffusion (equation \ref{PS.FN: eq: dPhidt final general tcf}).

The expression given by \textcite[eq.~42]{knobloch77} for the difference between the magnetic and scalar diffusivities can be simplified by assuming the velocity correlation is separable (equation \ref{PS: eq: separable correlator definition}).
However, the resulting expression contains an integral of the form $\int_0^t\d t_1 \int_0^{t_1} \d t_2 \int_0^{t_2} \d t_3 \, \tcf(t - t_1) \, \tcf(t_2 - t_3)$;
this integral diverges as $\bigO(t; t\to\infty)$.
It is unclear if this divergence is due to missed terms in their calculations, or a limitation of the cumulant expansion.
We note that the results reported by \textcite[eqs.~24,26]{nicklaus88} do not have this problem as the coefficient of this integral in their expressions turns out to be zero if the velocity field is Gaussian.

\subsubsection{Relation to the \texorpdfstring{$\alpha^2$}{alpha-squared} effect}
The negative contribution to the turbulent diffusivity in equation \ref{B.FN: emf O(tau) final general tcf} is reminiscent of that obtained by \textcite[eq.~4.8]{kraichnan1976diffusion} using a multi-scale averaging procedure.
However, in the calculations reported by \textcite{kraichnan1976diffusion}, the helicity of the velocity field does not have any effect on the mean passive scalar; he attributes this to differences between the conservation properties of the passive scalar and the magnetic field \parencite[p.~659]{kraichnan1976diffusion}.
This is contrary to our finding that helicity can even suppress the diffusion of the mean passive scalar (equation \ref{PS.FN: eq: dPhidt final general tcf}).
The fact that our $\bigO(\tau_c)$ correction to the turbulent diffusivity of the mean passive scalar becomes zero when the temporal correlation function is given by a top hat suggests that the results of \textcite{kraichnan1976diffusion} are attributable to his use of a `renovating flow' model (which corresponds to such a temporal correlation function).

\subsubsection{Comparison with renormalization group theory}

In simulations of forced turbulence, \citet[fig.~4]{BraSchRog17} found that the turbulent diffusivity is smaller when the turbulence is helically forced than when it is nonhelically forced.
Further, they found that for small $\Rm$, the helicity-dependent correction to the turbulent diffusivity of the magnetic field scales as $\Rm^2$.
This was later confirmed by \citet{Miz23} using renormalization group theory.
Since we have focused on the limit of high $\Rm$, we do not recover this scaling.

\Citet[eq.~18]{Miz23} also derived a correction to the turbulent diffusivity in the limit of small fractional helicity and large $\Rm$.
This correction seems consistent with our result (in the sense that the difference between the diffusivity in a helical velocity field and that in a nonhelical velocity field depends on the square of the helicity).
While we assume a single scale-independent correlation time, the method used by \citet{Miz23} allows the scale-dependent correlation time to be determined by the equations of motion (and thus it does not appear as a free parameter).
Note that \citet{Miz23} do not seem to have calculated the corrections to the $\alpha$ effect (see our equation \ref{B.FN: emf O(tau) final general tcf}); nevertheless, we expect such corrections to be obtainable using their method.

\section{Conclusions}
\label{passive-tensor: section: conclusions}

Conventional treatments of mean-field theory use the quasilinear approximation to derive expressions for transport coefficients which are exact when the correlation time of the velocity field is zero.
Assuming the velocity field is a Gaussian random field with a nonzero correlation time, we have applied the Furutsu-Novikov theorem to find the lowest-order corrections to the transport coefficients for two kinds of passive tensors.

For the diffusion of the mean passive scalar in the limit of high Peclet number, we have verified that our result matches earlier results obtained through different methods \citep{drummond1982, knobloch77}.
Using the multi-scale averaging approach, \textcite{kraichnan1976diffusion} has reported that the helicity of the velocity field does not affect the turbulent diffusion of the passive scalar.
We have found that this is only the case for a specific form of the temporal correlation function of the velocity field, corresponding to the renovating flow model used in that work.

We have also considered the mean magnetic field (treated, in the kinematic limit, as a passive pseudovector) at high magnetic Reynolds number, and
found that both the $\alpha$ effect and magnetic diffusion are affected by the correlation time of the velocity field.
An earlier result for the diffusivity of the mean magnetic field \parencite[eqs.~24,26]{nicklaus88} turns out to be identical to that for the mean passive scalar; we obtain extra (negative) contributions to the diffusivity of the mean magnetic field.

The validity of the expressions we have derived is limited by our assumption that the correlation time of the velocity field is scale-independent.
While the general formalism we have used can be adapted to account for a scale-dependent correlation time, this is left to future work.

\backsection[Acknowledgements]{
We thank Alexandra Elbakyan for facilitating access to scientific literature.
We acknowledge discussions with Matthias Rheinhardt, Igor Rogachevskii, Axel Brandenburg, and Petri K\"apyl\"a.
}

\backsection[Funding]{
This research received no specific grant from any funding agency, commercial or not-for-profit sectors.
}

\backsection[Software]{
Sympy \citep{sympy2017}.
}

\backsection[Declaration of interests]{
The authors report no conflict of interest.
}

\backsection[Author ORCID]{
GK, \url{https://orcid.org/0000-0003-2620-790X}; 
NS, \url{https://orcid.org/0000-0001-6097-688X}
}

\backsection[Author contributions]{
GK and NS conceptualized the research, interpreted the results, and wrote the paper.
GK performed the calculations.
}

\bibliographystyle{jfm}
\bibliography{refs.bib}

\begin{thebibliography}{30}
\expandafter\ifx\csname natexlab\endcsname\relax\def\natexlab#1{#1}\fi
\def\au#1{#1} \def\ed#1{#1} \def\yr#1{#1}\def\at#1{#1}\def\jt#1{\textit{#1}}
  \def\bt#1{#1}\def\bvol#1{\textbf{#1}} \def\vol#1{#1} \def\pg#1{#1}
  \def\publ#1{#1}\def\arxiv#1{#1}\def\org#1{#1}\def\st#1{\textit{#1}}

\bibitem[Brandenburg {\em et~al.\/}(2017)Brandenburg, Schober \&
  Rogachevskii]{BraSchRog17}
{\sc \au{Brandenburg, A.}, \au{Schober, J.} \& \au{Rogachevskii, I.}} \yr{2017}
   \at{The contribution of kinetic helicity to turbulent magnetic diffusivity}.
   \jt{Astronomische Nachrichten}  \bvol{338}~(7),  \pg{790–793}.

\bibitem[Brandenburg \& Subramanian(2005{\natexlab{{\em
  a\/}}})]{KanduPhysicsReports2005}
{\sc \au{Brandenburg, Axel} \& \au{Subramanian, Kandaswamy}}
  \yr{2005{\natexlab{{\em a\/}}}}  \at{Astrophysical magnetic fields and
  nonlinear dynamo theory}.  \jt{Physics Reports}  \bvol{417}~(1-4),
  \pg{1–209}.

\bibitem[Brandenburg \& Subramanian(2005{\natexlab{{\em b\/}}})]{BraSub05}
{\sc \au{Brandenburg, Axel} \& \au{Subramanian, K.}} \yr{2005{\natexlab{{\em
  b\/}}}}  \at{Minimal tau approximation and simulations of the alpha effect}.
  \jt{A\&A}  \bvol{439}~(3),  \pg{835–843}.

\bibitem[Davidson(2004)]{davidson2004turbulence}
{\sc \au{Davidson, P.~A.}} \yr{2004} {\em Turbulence: an introduction for
  scientists and engineers\/}.  \publ{Oxford Univ. Press}.

\bibitem[Drummond(1982)]{drummond1982}
{\sc \au{Drummond, I.~T.}} \yr{1982}  \at{Path-integral methods for turbulent
  diffusion}.  \jt{Journal of Fluid Mechanics}  \bvol{123},  \pg{59–68}.

\bibitem[Furutsu(1963)]{Fur63}
{\sc \au{Furutsu, K.}} \yr{1963}  \at{On the statistical theory of
  electromagnetic waves in a fluctuating medium (i)}.  \jt{Journal of Research
  of the National Bureau of Standards-D. Radio Propagation}  \bvol{67D}~(3),
  \pg{303–323}.

\bibitem[Gleeson(2000)]{gleeson2000closure}
{\sc \au{Gleeson, James~P.}} \yr{2000}  \at{A closure method for random
  advection of a passive scalar}.  \jt{Physics of Fluids}  \bvol{12}~(6),
  \pg{1472–1484}.

\bibitem[Gopalakrishnan \& Singh(2023)]{GopSin23}
{\sc \au{Gopalakrishnan, Kishore} \& \au{Singh, Nishant~K.}} \yr{2023}
  \at{Mean-field dynamo due to spatio-temporal fluctuations of the turbulent
  kinetic energy}.  \jt{Journal of Fluid Mechanics}  \bvol{973},  \pg{A29}.

\bibitem[Gopalakrishnan \& Singh(2024)]{GopSin24}
{\sc \au{Gopalakrishnan, Kishore} \& \au{Singh, Nishant~K}} \yr{2024}
  \at{Small-scale dynamo with nonzero correlation time}.  \jt{The Astrophysical
  Journal}  \bvol{970}~(1),  \pg{64}.

\bibitem[Gopalakrishnan \& Subramanian(2023)]{GopSub23}
{\sc \au{Gopalakrishnan, Kishore} \& \au{Subramanian, Kandaswamy}} \yr{2023}
  \at{Magnetic helicity fluxes from triple correlators}.  \jt{The Astrophysical
  Journal}  \bvol{943}~(1),  \pg{66}.

\bibitem[Jones(2011)]{jones11}
{\sc \au{Jones, Chris~A.}} \yr{2011}  \at{Planetary magnetic fields and fluid
  dynamos}.  \jt{Annual Review of Fluid Mechanics}  \bvol{43}~(1),
  \pg{583–614}.

\bibitem[Kearsley \& Fong(1975)]{KeaFon75}
{\sc \au{Kearsley, Elliot~A.} \& \au{Fong, Jeffrey~T.}} \yr{1975}  \at{Linearly
  independent sets of isotropic cartesian tensors of ranks up to eight}.
  \jt{Journal of Research of the National Bureau of Standards, Section B:
  Mathematical Sciences}  \bvol{79B}~(1–2),  \pg{49–58}.

\bibitem[Knobloch(1977)]{knobloch77}
{\sc \au{Knobloch, Edgar}} \yr{1977}  \at{The diffusion of scalar and vector
  fields by homogeneous stationary turbulence}.  \jt{Journal of Fluid
  Mechanics}  \bvol{83}~(1),  \pg{129–140}.

\bibitem[Kraichnan(1976)]{kraichnan1976diffusion}
{\sc \au{Kraichnan, Robert~H.}} \yr{1976}  \at{Diffusion of weak magnetic
  fields by isotropic turbulence}.  \jt{J. Fluid Mech}  \bvol{75}~(4),
  \pg{657–676}.

\bibitem[Krause \& Rädler(1980)]{KrauseRadler80}
{\sc \au{Krause, F.} \& \au{Rädler, K.-H.}} \yr{1980} {\em Mean-Field
  Magnetohydrodynamics and Dynamo Theory\/}, 1st edn.  \publ{Pergamon press}.

\bibitem[Käpylä {\em et~al.\/}(2006)Käpylä, {Korpi, M. J.}, {Ossendrijver}
  \& {Tuominen}]{kapyla2006strouhal}
{\sc \au{Käpylä, Petri~J.}, \au{{Korpi, M. J.}}, \au{{Ossendrijver}, M.} \&
  \au{{Tuominen}, I.}} \yr{2006}  \at{Local models of stellar convection.
  {III}. the {S}trouhal number}.  \jt{\aap}  \bvol{448}~(2),  \pg{433–438}.

\bibitem[Lesieur(2008)]{lesieur2008turbulence}
{\sc \au{Lesieur, Marcel}} \yr{2008} {\em Turbulence in Fluids\/}, 4th edn.,
  \st{Fluid Mechanics and its Applications},  \vol{vol.~84}.  \publ{Springer}.

\bibitem[Meurer {\em et~al.\/}(2017)Meurer, Smith, Paprocki, Čertík,
  Kirpichev, Rocklin, Kumar, Ivanov, Moore, Singh, Rathnayake, Vig, Granger,
  Muller, Bonazzi, Gupta, Vats, Johansson, Pedregosa, Curry, Terrel, Roučka,
  Saboo, Fernando, Kulal, Cimrman \& Scopatz]{sympy2017}
{\sc \au{Meurer, Aaron}, \au{Smith, Christopher~P.}, \au{Paprocki, Mateusz},
  \au{Čertík, Ondřej}, \au{Kirpichev, Sergey~B.}, \au{Rocklin, Matthew},
  \au{Kumar, AMiT}, \au{Ivanov, Sergiu}, \au{Moore, Jason~K.}, \au{Singh,
  Sartaj}, \au{Rathnayake, Thilina}, \au{Vig, Sean}, \au{Granger, Brian~E.},
  \au{Muller, Richard~P.}, \au{Bonazzi, Francesco}, \au{Gupta, Harsh},
  \au{Vats, Shivam}, \au{Johansson, Fredrik}, \au{Pedregosa, Fabian},
  \au{Curry, Matthew~J.}, \au{Terrel, Andy~R.}, \au{Roučka, Štěpán},
  \au{Saboo, Ashutosh}, \au{Fernando, Isuru}, \au{Kulal, Sumith}, \au{Cimrman,
  Robert} \& \au{Scopatz, Anthony}} \yr{2017}  \at{{S}ym{P}y: symbolic
  computing in {P}ython}.  \jt{PeerJ Computer Science}  \bvol{3},  \pg{e103}.

\bibitem[Mizerski(2023)]{Miz23}
{\sc \au{Mizerski, Krzysztof~A.}} \yr{2023}  \at{Helical correction to
  turbulent magnetic diffusivity}.  \jt{Phys. Rev. E}  \bvol{107},
  \pg{055205}.

\bibitem[Moffatt(1978)]{MoffattMagFieldGenBook}
{\sc \au{Moffatt, Henry~Keith}} \yr{1978} {\em Magnetic Field Generation In
  Electrically Conducting Fluids\/}.  \publ{Cambridge University Press}.

\bibitem[Monin \& Yaglom(1971)]{MoninYaglomVol1}
{\sc \au{Monin, A.~S.} \& \au{Yaglom, A.~M.}} \yr{1971} {\em Statistical Fluid
  Mechanics: Mechanics of Turbulence\/}, ,  \vol{vol.~1}.  \publ{The MIT
  Press}.

\bibitem[Nicklaus \& Stix(1988)]{nicklaus88}
{\sc \au{Nicklaus, Bernhard} \& \au{Stix, Michael}} \yr{1988}  \at{Corrections
  to first order smoothing in mean-field electrodynamics}.  \jt{Geophysical \&
  Astrophysical Fluid Dynamics}  \bvol{43}~(2),  \pg{149–166}.

\bibitem[Novikov(1965)]{novikov1965}
{\sc \au{Novikov, Evgenii~A.}} \yr{1965}  \at{Functionals and the random-force
  method in turbulence theory}.  \jt{Sov. Phys. JETP}  \bvol{20}~(5),
  \pg{1290–1294}.

\bibitem[Roberts \& Soward(1975)]{robertsSoward75}
{\sc \au{Roberts, P.~H.} \& \au{Soward, A.~M.}} \yr{1975}  \at{A unified
  approach to mean field electrodynamics}.  \jt{Astronomische Nachrichten}
  \bvol{296}~(2),  \pg{49–64}.

\bibitem[Schekochihin \& Kulsrud(2001)]{SchekochihinKulsrud2001}
{\sc \au{Schekochihin, Alexander~A.} \& \au{Kulsrud, Russell~M.}} \yr{2001}
  \at{Finite-correlation-time effects in the kinematic dynamo problem}.
  \jt{Physics of Plasmas}  \bvol{8}~(11),  \pg{4937–4953}.

\bibitem[Shukurov \& Subramanian(2022)]{ShukurovKanduBook}
{\sc \au{Shukurov, Anvar} \& \au{Subramanian, Kandaswamy}} \yr{2022} {\em
  Astrophysical Magnetic Fields: From Galaxies to the Early Universe\/}. {\em
  Cambridge Astrophysics\/} 56.  \publ{Cambridge University Press}.

\bibitem[Silant'ev(2000)]{silantev2000}
{\sc \au{Silant'ev, N.~A.}} \yr{2000}  \at{Magnetic dynamo due to turbulent
  helicity fluctuations}.  \jt{\aap}  \bvol{364},  \pg{339–347}.

\bibitem[Singh(2016)]{singh2016moffatt}
{\sc \au{Singh, Nishant~Kumar}} \yr{2016}  \at{Moffatt-drift-driven large-scale
  dynamo due to $\alpha$ fluctuations with non-zero correlation times}.
  \jt{Journal of Fluid Mechanics}  \bvol{798},  \pg{696–716}.

\bibitem[White(1999)]{Whi99}
{\sc \au{White, Frank~M.}} \yr{1999} {\em Fluid Mechanics\/}, 4th edn.
  \publ{WCB/McGraw-Hill}.

\bibitem[Zhang \& Mahajan(2017)]{ZhangMahajan2017}
{\sc \au{Zhang, Y.~Z.} \& \au{Mahajan, S.~M.}} \yr{2017}  \at{Limitations of
  the clump-correlation theories of shear-induced turbulence suppression}.
  \jt{Physics of Plasmas}  \bvol{24}~(5),  \pg{054502}.

\end{thebibliography}

\appendix
\allowdisplaybreaks

\section{\texorpdfstring{$n$}{n}-th functional derivative of \texorpdfstring{$\nabla\theta$}{grad(theta)}}
\label{PS: appendix: nth functional derivative recursion-like relation}
We first recall the definition of the $n$-th functional derivative of a functional $F[f]$ wrt. a function $f(t)$:
\begin{equation}
	\left. \frac{\d^n F[f+\epsilon\chi]}{\d\epsilon^n} \right|_{\epsilon=0} = \int \frac{\delta^n F}{\delta f(t_1)\dots\delta f(t_n)} \chi(t_1)\dots\chi(t_n) \,\d t_1\dots\d t_n
	\label{PS: eq: functional derivative definition}
\end{equation}
where $\epsilon$ is a real number, and $\chi(t)$ is an arbitrary test function.\footnote{
One might be confused about why there is no $1/n!$ or similar term on the RHS, but one can confirm this is correct by considering a functional $\int A(x_1,x_2) \, \d x_1\d x_2$ and comparing it with the Taylor series expansion for a functional \parencite[e.g.][eq.~2.4]{novikov1965}.
}

Using the same notation as in section \ref{PS: section: FuruNovi corrections to FOSA}, we write (similar to equation \ref{PS: eq: implicit integral equation for lambda})
\begin{equation}
	\lambda_i(\vec{x},t) = - \frac{\dh}{\dh x_i} \int_{-\infty}^t\d\tau \int\d \vec{q} \, G(\vec{x},t|\vec{q}, \tau) \, u_m(\vec{q},\tau)\lambda_m(\vec{q},\tau)
	\,.
\end{equation}
Denoting $\varied{\lambda} \defn \lambda[u_m+\epsilon\chi_m]$, we write
\begin{equation}
	\varied{\lambda}_i(\vec{x},t) = - \frac{\dh}{\dh x_i} \int_{-\infty}^t\d\tau \int\d \vec{q} \, G(\vec{x},t|\vec{q}, \tau) \left( u_m(\vec{q},\tau) + \epsilon\chi_m(\vec{q},\tau) \right) \varied{\lambda}_m(\vec{q},\tau)
	\,.
\end{equation}
Taking $n$ derivatives on both sides and setting $\epsilon=0$, we write
\begin{align}
	\begin{split}
		\left. \frac{\d^n \varied{\lambda}_i(\vec{x},t)}{\d \epsilon^n} \right|_{\epsilon=0} ={}& - \frac{\dh}{\dh x_i} \int_{-\infty}^t\d\tau \int\d \vec{q} \, G(\vec{x},t|\vec{q}, \tau) \,u_m(\vec{q},\tau) \left. \frac{\d^{n} \varied{\lambda}_m(\vec{q},\tau) }{\d\epsilon^{n}} \right|_{\epsilon=0}
		\\& - n \frac{\dh}{\dh x_i} \int_{-\infty}^t\d\tau \int\d \vec{q} \, G(\vec{x},t|\vec{q}, \tau)  \,\chi_m(\vec{q},\tau) \left. \frac{\d^{n-1} \varied{\lambda}_m(\vec{q},\tau) }{\d\epsilon^{n-1}} \right|_{\epsilon=0}
	\end{split}
\end{align}
which gives us the following relation between the functional derivatives (we use the notation $u_{j_i} \defn u_{j_{i}}(\vec{x}^{(i)},t^{(i)})$):
\begin{align}
	\begin{split}
		\frac{\delta^n \lambda_i(\vec{x},t)}{\delta u_{j_1}\dots \delta u_{j_n}}
		={}& - \frac{\dh}{\dh x_i} \int\d \vec{q} \int_{\op{max}(t^{(1)}, \dots , t^{(n)})}^t \hspace{-4.5em} \d\tau \, G(\vec{x},t|\vec{q}, \tau)\, u_m(\vec{q},\tau) \,\frac{\delta^n \lambda_m(\vec{q},\tau)}{\delta u_{j_1}\dots \delta u_{j_n}}
		\\& - \sum_{\alpha=1}^n \frac{\dh}{\dh x_i} G(\vec{x},t|\vec{x}^{(\alpha)}, t^{(\alpha)})  \Heaviside(t-t^{(\alpha)}) \, \frac{\delta^{n-1} \lambda_{j_\alpha}(\vec{x}^{(\alpha)},t^{(\alpha)})}{\delta u_{j_1}\dots \delta u_{j_{\alpha-1}} \delta u_{j_{\alpha+1}} \dots\delta u_{j_n}}
		\,.
	\end{split}
\end{align}
We average both sides of the above and use the Furutsu-Novikov theorem (equation \ref{PS: eq: use FN theorem to write uGradphi term}) to write
\begin{align}
	\begin{split}
		&\meanBr{ \frac{\delta^n \lambda_i(\vec{x},t)}{\delta u_{j_1} \dots \delta u_{j_n}} }
		\\&= - \frac{\dh}{\dh x_i} \int\d \vec{q}\d\vec{x}'\d t' \int_{\op{max}(t^{(1)}, \dots , t^{(n)})}^t \hspace{-4.5em} \d\tau \, G(\vec{x},t|\vec{q}, \tau) \meanBr{ u_m(\vec{q},\tau) u_a(\vec{x}',t') } \meanBr{ \frac{\delta^{n+1} \lambda_m(\vec{q},\tau)}{\delta u_a(\vec{x}',t') \delta u_{j_1}\dots \delta u_{j_n} } }
		\\&\phantom{={}} - \sum_{\alpha=1}^n \frac{\dh}{\dh x_i} G(\vec{x},t|\vec{x}^{(\alpha)}, t^{(\alpha)})  \Heaviside(t-t^{(\alpha)}) \meanBr{ \frac{\delta^{n-1} \lambda_{j_\alpha}(\vec{x}^{(\alpha)},t^{(\alpha)})}{\delta u_{j_1}\dots \delta u_{j_{\alpha-1}} \delta u_{j_{\alpha+1}} \dots\delta u_{j_n}} }
		\,.
	\end{split}
\end{align}
By taking causality into account ($\delta\lambda(t)/\delta u(t') = 0$ if $t'>t$) we can write the above more compactly as
\begin{align}
	\begin{split}
		&\meanBr{ \frac{\delta^n \lambda_i(\vec{x},t)}{\delta u_{j_1} \dots \delta u_{j_n}} }
		\\&= - \frac{\dh}{\dh x_i} \int\d \vec{q}\d\vec{x}'\d t' \d\tau \, G^c\!(\vec{x},t|\vec{q}, \tau) \meanBr{ u_m(\vec{q},\tau) u_a(\vec{x}',t') } \meanBr{ \frac{\delta^{n+1} \lambda_m(\vec{q},\tau)}{\delta u_a(\vec{x}',t') \delta u_{j_1}\dots \delta u_{j_n} } }
		\\&\phantom{={}} - \sum_{\alpha=1}^n \frac{\dh}{\dh x_i} G^c\!(\vec{x},t|\vec{x}^{(\alpha)}, t^{(\alpha)}) \meanBr{ \frac{\delta^{n-1} \lambda_{j_\alpha}(\vec{x}^{(\alpha)},t^{(\alpha)})}{\delta u_{j_1}\dots \delta u_{j_{\alpha-1}} \delta u_{j_{\alpha+1}} \dots\delta u_{j_n}} }
		\,.
	\end{split} \label{PS: eq: nth functional derivative recursion-like relation}%
\end{align}
Note the superscript `$c$' on the Green function, which denotes that it has been made causal ($G^{c}\!(x,t|x',t') = 0$ if $t'>t$; $G(x,t|x',t')$ otherwise).

\section{Coefficients that depend on the temporal correlation function}
\label{PS: appendix: g1 g2 defn properties}

Our results depend on the form of the temporal correlation function of the velocity field only through the following coefficients:
\begin{subequations}
	\begin{align}
		\begin{split}
			g_1
			&{}\defn
			\frac{1}{\tau_c} \int_{-\infty}^t \d t' \int_{-\infty}^{ t' } \d t_1 \int_{-\infty}^{ t_1 } \d t_2 \, \tcf^{(t-t_1)} \tcf^{(t'-t_2)}
		\end{split}
		\\
		\begin{split}
			g_2
			&{}\defn
			\frac{1}{\tau_c} \int_{-\infty}^t \d t' \int_{-\infty}^{ t' } \d t_2 \int_{-\infty}^{ t_2 } \d t_1 \, \tcf^{(t-t_1)} \tcf^{(t'-t_2)}
			\,.
		\end{split}
	\end{align} \label{PS: eq: g1 g2 defn}%
\end{subequations}
The values of these coefficients depend on the form of the temporal correlation function (table \ref{PS: table: g2 values}).
Regardless of the form of the temporal correlation function, they satisfy the identity \parencite[appendix C]{GopSin24}
\begin{equation}
	g_1 + g_2 = \frac{1}{4}
	\,.
	\label{B2.SSD.real: eq: g1 + g2}
\end{equation}

\section{Velocity correlations for locally isotropic, weakly inhomogeneous turbulence}
\sectionmark{Weakly inhomogeneous velocity correlations}
\label{PS: appendix: weakly inhomogeneous correlators}

\subsection{The equal-time correlation in Fourier-space}
\label{PS: appendix: derive RobertsSoward expressions}

We define the Fourier transform as
\begin{equation}
	\tilde{f}(\vec{k}) = \int\frac{\d^3x}{(2\pi)^3}\, e^{-i \vec{k}\cdot\vec{x}} f(\vec{x}) \label{PS: eq: fourier transform definition deriving Roberts Soward}
\end{equation}
and the two-point correlation of the Fourier transformed velocity field as $V_{ij}(\vec{k},\vec{K},t) \defn \meanBr{ \tilde{u}_i(\tfrac{1}{2}\vec{K}+\vec{k},t) \tilde{u}_j(\tfrac{1}{2}\vec{K}-\vec{k},t) }$.
If $\vec{k}_1$ and $\vec{k}_2$ denote the Fourier conjugates of the two positions between which the correlation is taken, $\vec{K} \defn \vec{k}_1 + \vec{k}_2$, and $\vec{k} \defn ( \vec{k}_1-\vec{k}_2 )/2$.
We interpret $\vec{K}$ as the `large scale' wavevector, and $\vec{k}$ as the `small scale' wavevector.
According to the notion of weak inhomogeneity introduced by \textcite{robertsSoward75}, one
Taylor-expands this double correlation function function as a series in $\vec{K}$;
assumes the lowest-order ($\vec{K}$-independent) terms are identical to those for homogeneous and isotropic turbulence (\citealp[eq.~7.56]{MoffattMagFieldGenBook}; \citealp[eq.~5.84]{lesieur2008turbulence});
assumes that the higher-order terms only depend on the energy and helicity spectra (along with $\vec{k}$ and $\vec{K}$); and
discards $\bigO(K^2)$ terms.
Requiring the velocity field to be incompressible then leads to
\begin{align}
	\begin{split}
		V_{ij}(\vec{k},\vec{K},t) ={}& \op{P}_{ij}(\vec{k})E(k,\vec{K},t) - \frac{i}{k^2}\epsilon_{ijc}k_cF(k,\vec{K},t) 
		\\& + \frac{1}{2k^2} \left( k_j \delta_{ik} - k_i \delta_{jk} \right) K_k E(k,\vec{K},t) 
		\\& + \frac{i}{2k^4} \left( k_i \epsilon_{jck} + k_j \epsilon_{ick} \right) k_c K_k F(k,\vec{K},t) + \bigO(K^2)
		\,.
	\end{split} \label{PS: eq: vij fourier}
\end{align}

\subsection{Equal-time single-point correlations in real space}
\label{PS: appendix: equal-time correlators real space}

\subsubsection{An example}
As an example, we demonstrate how one can use equation \ref{PS: eq: vij fourier} to derive an expression for $\meanBr{u_iu_j}$ (an equal-time single-point correlation).
We start from
\begin{equation}
	\meanBr{u_i(\vec{x})u_j(\vec{x}) }
	=
	\int\d^3p\d^3q\, e^{i\left(\vec{p}+\vec{q}\right)\cdot\vec{x}} \meanBr{\tilde{u}_i(\vec{q})\tilde{u}_j(\vec{p}) }
	\,.
\end{equation}
Defining $\vec{K} \defn \vec{p}+\vec{q}$ and $\vec{k} \defn \frac{1}{2}\left(\vec{p}-\vec{q}\right)$,
\begin{equation}
	\meanBr{ u_i(\vec{x})u_j(\vec{x}) }
	=
	\int\d^3K\d^3k\, e^{i\vec{K}\cdot\vec{x}} \meanBr{ \tilde{u}_i(\tfrac{1}{2}\vec{K}-\vec{k}) \tilde{u}_j(\tfrac{1}{2}\vec{K}+\vec{k}) }
	\,.
\end{equation}
Using equation \ref{PS: eq: vij fourier},
\begin{align}
	\begin{split}
		\meanBr{ u_i(\vec{x})u_j(\vec{x}) }
		={}&
		\int\d^3K k^2\d k\d\Omega_k\, e^{i\vec{K}\cdot\vec{x}} 
		\bigg(
		\op{P}_{ji}(\vec{k})E(k,\vec{K},t) - \frac{i}{k^2}\epsilon_{jic}k_cF(k,\vec{K},t) 
		\\& - \frac{1}{2k^2}k_jK_iE(k,\vec{K},t) 
		+ \frac{1}{2k^2}k_iK_jE(k,\vec{K},t)
		\\& + \frac{i}{2k^4}k_j\epsilon_{icd}k_cK_dF(k,\vec{K},t)
		+\frac{i}{2k^4}k_i\epsilon_{jcd}k_cK_dF(k,\vec{K},t)
		\bigg)
		\,.
	\end{split} \label{PS: eq: uiuj single-point with unevaluated angular integral}%
\end{align}
Angular integrals over products of $\vec{k}$ can be evaluated by noting that the result has to be an isotropic tensor (and so needs to be constructed using $\delta_{ij}$ and $\epsilon_{ijk}$, \citealp[e.g.][]{KeaFon75}) and must be consistent with the symmetries of the integrand:
\begin{align}
	\int\d\Omega \, k_i ={}& 0
	\\
	\int\d\Omega \, k_i k_j ={}& \frac{4\pi k^2}{3} \delta_{ij}
	\\
	\int\d\Omega \, k_i k_j k_m ={}& 0
	\\
	\int\d\Omega \, k_i k_j k_m k_n ={}& \frac{4\pi k^4}{15} \left( \delta_{ij}\delta_{mn} + \delta_{im}\delta_{jn} + \delta_{in}\delta_{jm} \right)
	\,.
\end{align}
Using these, equation \ref{PS: eq: uiuj single-point with unevaluated angular integral} can be written as
\begin{align}
	\begin{split}
		\meanBr{ u_i(\vec{x})u_j(\vec{x}) }
		={}&
		\frac{1}{3}\delta_{ji}\meanBr{u^2(\vec{x},t)}
		\,.
	\end{split} \label{PS: eq: vivj single-point correlation}
\end{align}
The steps to derive the other expressions that follow are similar to those above, so we shall omit the intermediate steps and just state the results.

\subsubsection{Results}
\begin{align}
	\begin{split}
		\meanBr{ u_i u_j }
		={}&
		\frac{1}{3}\delta_{ji}\meanBr{u^2}
		+ \bigO(\dh^2)
	\end{split}
	\\
	\begin{split}
		\meanBr{ u_i \frac{\dh u_j }{\dh x_k} } 
		={}& 
		\left[ \frac{1}{6} \delta_{ij} \dh_k + \frac{1}{12} \left( \delta_{ik} \dh_j - \delta_{jk} \dh_i \right) \right] \meanBr{u^2}
		- \frac{1}{6} \epsilon_{ijk} h
		+ \bigO(\dh^2)
	\end{split}
	\\
	\begin{split}
		\meanBr{ u_i \dh_a\dh_b u_j } 
		={}& 
		\frac{1}{12} \epsilon_{jia} \dh_b h
		+ \frac{1}{12} \epsilon_{jib} \dh_a h
		+ \left(  \frac{1}{30}\left[\delta_{ai}\delta_{bj} + \delta_{aj}\delta_{bi}\right] - \frac{2}{15}\delta_{ab}\delta_{ij} \right) \meanBr{\omega^2}
		\\& 
		- \frac{1}{60} \left( 2\delta_{ab}\epsilon_{ijk} + \delta_{jb}\epsilon_{iak} + \delta_{aj}\epsilon_{ibk} + \delta_{ib}\epsilon_{jak} + \delta_{ai}\epsilon_{jbk} \right) \dh_k h
		+ \bigO(\dh^2)
		\label{PS: eq: <v_id_ad_bv_j>}
	\end{split}
	\\
	\begin{split}
		\meanBr{ u_i \dh_a\nabla^2 u_j } 
		={}& 
		\left( - \frac{1}{20}\delta_{ai}\dh_j + \frac{7}{60}\delta_{aj}\dh_i - \frac{3}{10}\delta_{ij}\dh_a \right) \meanBr{\omega^2}
		- \frac{1}{6} \epsilon_{jia} \mathcal{L}
		+ \bigO(\dh^2)
		\label{PS: eq: <v_id_ad^2v_j>}
	\end{split}
\end{align}
where $\vec{\omega} \defn \curl\vec{u}$;
$h \defn \meanBr{\vec{u}\cdot\vec{\omega}}$; and
$\mathcal{L} \defn \meanBr{\vec{\omega} \cdot ( \curl\vec{\omega} ) }$.
We have also used the relations $\meanBr{\omega^2}(\vec{R},t) = \int 8\pi k^4 E(k, \vec{R}, t) \,\d k$ \parencite[eq.~7.51]{MoffattMagFieldGenBook}; and $\mathcal{L}(\vec{R},t) = \int \d k\, 8\pi k^4 F(k, \vec{R}, t) $.
Aside, we note that the expression given by \textcite[eq.~A16]{GopSub23} for $\meanBr{ u_i \dh_a\nabla^2 u_j }$ is wrong.\footnote{
This can be checked by setting $i=a$ in equation \ref{PS: eq: <v_id_ad^2v_j>} and comparing it with $\dh_i \meanBr{u_i \Lap u_j}$ calculated using equation \ref{PS: eq: <v_id_ad_bv_j>} \parencite[which is the same as eq.~A13 of][]{GopSub23}.
}

\subsection{Generalization to unequal-time correlations}
\label{PS: appendix: unequal-time-correlators separable}
Assuming the velocity correlations are separable (equation \ref{PS: eq: separable correlator definition}), the results in appendix \ref{PS: appendix: equal-time correlators real space} can be easily generalized as
\begin{align}
	\begin{split}
		C_{ij}
		={}&
		\frac{2}{3}\delta_{ij} E
	\end{split}
	\\
	\begin{split}
		C_{ijk}
		={}&
		\frac{1}{6} \epsilon_{ikj} H
		+ \left[ \frac{1}{3} \delta_{ij} \dh_k + \frac{1}{6} \left( \delta_{ik} \dh_j - \delta_{jk} \dh_i \right) \right] E
	\end{split}
	\\
	\begin{split}
		C_{ijkk}
		={}&
		- \frac{1}{3} \delta_{ij} N
		- \frac{1}{6} \epsilon_{ija} \dh_a H
	\end{split}
	\\
	\begin{split}
		C_{ijab}
		={}&
		\left(  \frac{1}{30}\left[\delta_{ai}\delta_{bj} + \delta_{aj}\delta_{bi}\right] - \frac{2}{15}\delta_{ab}\delta_{ij} \right) N
		+ \frac{1}{12} \epsilon_{jia} \dh_b H
		+ \frac{1}{12} \epsilon_{jib} \dh_a H
		\\&
		- \frac{1}{60} \left( 2\delta_{ab}\epsilon_{ijk} + \delta_{jb}\epsilon_{iak} + \delta_{aj}\epsilon_{ibk} + \delta_{ib}\epsilon_{jak} + \delta_{ai}\epsilon_{jbk} \right) \dh_k H
	\end{split}
	\\
	\begin{split}
		C_{ijabb}
		={}&
		- \frac{1}{6} \epsilon_{jia} L
		+ \left( - \frac{1}{20}\delta_{ai}\dh_j + \frac{7}{60}\delta_{aj}\dh_i - \frac{3}{10}\delta_{ij}\dh_a \right) N
	\end{split}
\end{align}
where $E$, $H$, and $N$ are defined in equations \ref{PS: eq: E H N definitions}, and $L$ is defined in equation \ref{PS: eq: L definition}.

\section{Simplification of the \texorpdfstring{$\bigO(\tau_c)$}{O(tau)} corrections to the evolution equation for the mean passive scalar}
\sectionmark{\texorpdfstring{$\bigO(\tau_c)$}{O(tau)} corrections for mean passive scalar}
\label{PS: appendix: simplify v4 terms alt method}

We now simplify the terms on the RHS of equation \ref{PS: eq: series with terms up to v4}.

\subsection{The first higher-order term}
Let us assume $\kappa=0$ (this corresponds to neglecting $\bigO(\kappa \tau_c)$ terms in the final evolution equation for $\Phi$).
We then have $G^c(\vec{x},t|\vec{q},\tau) = \delta(\vec{x}-\vec{q}) \Heaviside(t-\tau)$.
For convenience, we will use superscripts on the velocity fields to denote which ones are connected by averaging, such that velocity variables with matching superscripts are connected by averaging (e.g.\@ $\meanBr{uu} \to u^{(1)}u^{(1)}$).
Denoting $\funnyInt\dots \defn \int\d\tau_1\d\tau_2\d\tau_3 \Heaviside(\tau_1-\tau_3) \Heaviside(t-\tau_2) \Heaviside(\tau_2-\tau_1) \,\dots $;
assuming the velocity field is incompressible;
and integrating by parts as required, we write
\begin{align}
	\begin{split}
		\continuedTerm&\negphantom{\continuedTerm}
		- \velCorr{{\posTimeCombFree}_{i} {1}_{i_{1}}} \derG{i}{\posTimeCombFree|2} \velCorr{{2}_{i_{2}} {3}_{i_{3}}} \derG{i_{2}}{2|1} \derG{i_{1}}{1|3} \meanBr{\lambda_{i_{3}}\posTimeComb{3} }
		\\
		={}& - \velCorr{{\posTimeCombFree}_{i} {1}_{i_{1}}} \ddxDummy{\posTimeCombFree}_{i} G^c(\posTimeCombFree|2) \velCorr{{2}_{i_{2}} {3}_{i_{3}}} \ddxDummy{2}_{i_{2}}G^c(2|1) \ddxDummy{1}_{i_{1}} G^c(1|3) \meanBr{\lambda_{i_{3}}\posTimeComb{3} }
	\end{split}
	\\\begin{split}
		={}& - \funnyInt\int\d\vec{q}^{(1)} \d\vec{q}^{(2)} \d\vec{q}^{(3)} \bigg[ 
		\meanBr{u_i(\vec{x},t) u_{i_{1}}(\vec{q}^{(1)},\tau_1)} 
		\frac{\dh \delta(\vec{x}-\vec{q}^{(2)}) }{\dh x_i}  
		\\& \continuedTerm\times
		\meanBr{ u_{i_{2}}(\vec{q}^{(2)},\tau_2) u_{i_{3}}(\vec{q}^{(3)},\tau_3)} 
		\frac{\dh \delta(\vec{q}^{(2)}-\vec{q}^{(1)}) }{\dh q^{(2)}_{i_{2}}} 
		\frac{\dh \delta(\vec{q}^{(1)}-\vec{q}^{(3)}) }{\dh q^{(1)}_{i_{1}}} 
		\\& \continuedTerm\times
		\meanBr{\lambda_{i_{3}}(\vec{q}^{(3)}, \tau_3) }
		\bigg]
	\end{split}
	\\\begin{split}
		={}& - \frac{\dh}{\dh x_i} \funnyInt
		u_i^{(1)}(\vec{x},t) 
		\frac{\dh }{\dh x_{i_{2}}}
		\left(
		u_{i_{1}}^{(1)}(\vec{x},\tau_1) 
		u_{i_{2}}^{(2)}(\vec{x},\tau_2) 
		\frac{\dh  }{\dh x_{i_{1}}} 
		\left[
		u_{i_{3}}^{(2)}(\vec{x},\tau_3) 
		\meanBr{\lambda_{i_{3}}(\vec{x}, \tau_3) }
		\right]
		\right)
	\end{split}
	\\\begin{split}
		={}& - \frac{\dh}{\dh x_i} \funnyInt \meanBr{ u_i(\vec{x},t) \frac{\dh u_{i_{1}}(\vec{x},\tau_1) }{\dh x_{i_{2}}} } \meanBr{ u_{i_{2}}(\vec{x},\tau_2) \frac{\dh u_{i_{3}}(\vec{x},\tau_3) }{\dh x_{i_{1}}} } \meanBr{\lambda_{i_{3}}(\vec{x}, \tau_3) }
		\\& - \frac{\dh}{\dh x_i} \funnyInt \meanBr{ u_i(\vec{x},t)  u_{i_{1}}(\vec{x},\tau_1) } \meanBr{ u_{i_{2}}(\vec{x},\tau_2) \frac{\dh^2 u_{i_{3}}(\vec{x},\tau_3) }{\dh x_{i_{1}} \dh x_{i_{2}} } } \meanBr{\lambda_{i_{3}}(\vec{x}, \tau_3) }
		\\& - \frac{\dh}{\dh x_i} \funnyInt \meanBr{ u_i(\vec{x},t)  u_{i_{1}}(\vec{x},\tau_1) } \meanBr{ u_{i_{2}}(\vec{x},\tau_2) \frac{\dh u_{i_{3}}(\vec{x},\tau_3) }{\dh x_{i_{1}}} } \frac{\dh \meanBr{\lambda_{i_{3}}(\vec{x}, \tau_3) } }{\dh x_{i_{2}}}
		\\& - \frac{\dh}{\dh x_i} \funnyInt \meanBr{ u_i(\vec{x},t)  \frac{\dh u_{i_{1}}(\vec{x},\tau_1) }{\dh x_{i_{2}}} } \meanBr{ u_{i_{2}}(\vec{x},\tau_2) u_{i_{3}}(\vec{x},\tau_3) } \frac{\dh \meanBr{\lambda_{i_{3}}(\vec{x}, \tau_3) } }{\dh x_{i_{1}}} 
		\\& - \frac{\dh}{\dh x_i} \funnyInt \meanBr{ u_i(\vec{x},t)  u_{i_{1}}(\vec{x},\tau_1) } \meanBr{ u_{i_{2}}(\vec{x},\tau_2) \frac{\dh u_{i_{3}}(\vec{x},\tau_3) }{\dh x_{i_{2}}} } \frac{\dh \meanBr{\lambda_{i_{3}}(\vec{x}, \tau_3) } }{\dh x_{i_{1}}}
		\\& - \frac{\dh}{\dh x_i} \funnyInt \meanBr{ u_i(\vec{x},t) u_{i_{1}}(\vec{x},\tau_1) } \meanBr{ u_{i_{2}}(\vec{x},\tau_2) u_{i_{3}}(\vec{x},\tau_3) } \frac{\dh^2 \meanBr{\lambda_{i_{3}}(\vec{x}, \tau_3) } }{\dh x_{i_{1}} \dh x_{i_{2}} } 
		\,.
	\end{split} \label{PS: eq: Fn series thrice first extra term general}
\end{align}

\subsection{The second higher-order term}
Denoting $\funnyInt \dots \defn \int\d\tau_1\d\tau_2\d\tau_3 \Heaviside(t-\tau_2) \Heaviside(\tau_2-\tau_3) \Heaviside(\tau_3-\tau_1) \dots$ and following similar steps, we simplify the other term as follows.
\begin{align}
	\begin{split}
		\continuedTerm&\negphantom{\continuedTerm}
		- \velCorr{{\posTimeCombFree}_{i} {1}_{i_{1}}} \derG{i}{\posTimeCombFree|2} \velCorr{{2}_{i_{2}} {3}_{i_{3}}} \derG{i_{2}}{2|3} \derG{i_{3}}{3|1} \meanBr{\lambda_{i_{1}}\posTimeComb{1} }
		\\
		={}& - \frac{\dh}{\dh x_i} \funnyInt \meanBr{ u_i(\vec{x},t) \frac{\dh u_{i_{1}}(\vec{x},\tau_1) }{\dh x_{i_{3}}} } \meanBr{ u_{i_{2}}(\vec{x},\tau_2) \frac{\dh u_{i_{3}}(\vec{x},\tau_3) }{\dh x_{i_{2}}} } \meanBr{\lambda_{i_{1}}(\vec{x}, \tau_1) } 
		\\& - \frac{\dh}{\dh x_i} \funnyInt \meanBr{ u_{i_{2}}(\vec{x},\tau_2) u_{i_{3}}(\vec{x},\tau_3) } \meanBr{ u_i(\vec{x},t) \frac{\dh^2 u_{i_{1}}(\vec{x},\tau_1) }{\dh x_{i_{3}} \dh x_{i_{2}} } } \meanBr{\lambda_{i_{1}}(\vec{x}, \tau_1) }
		\\& - \frac{\dh}{\dh x_i} \funnyInt \meanBr{ u_{i_{2}}(\vec{x},\tau_2) u_{i_{3}}(\vec{x},\tau_3) } \meanBr{ u_i(\vec{x},t) \frac{\dh u_{i_{1}}(\vec{x},\tau_1) }{\dh x_{i_{3}}} } \frac{\dh \meanBr{\lambda_{i_{1}}(\vec{x}, \tau_1) } }{\dh x_{i_{2}}} 
		\\& - \frac{\dh}{\dh x_i} \funnyInt \meanBr{ u_{i_{2}}(\vec{x},\tau_2)  \frac{\dh u_{i_{3}}(\vec{x},\tau_3) }{\dh x_{i_{2}}} } \meanBr{ u_i(\vec{x},t) u_{i_{1}}(\vec{x},\tau_1) } \frac{\dh \meanBr{\lambda_{i_{1}}(\vec{x}, \tau_1) } }{\dh x_{i_{3}}}
		\\& - \frac{\dh}{\dh x_i} \funnyInt \meanBr{ u_{i_{2}}(\vec{x},\tau_2) u_{i_{3}}(\vec{x},\tau_3) } \meanBr{ u_i(\vec{x},t) \frac{\dh u_{i_{1}}(\vec{x},\tau_1) }{\dh x_{i_{2}}} } \frac{\dh \meanBr{\lambda_{i_{1}}(\vec{x}, \tau_1) } }{\dh x_{i_{3}}} 
		\\& - \frac{\dh}{\dh x_i} \funnyInt \meanBr{ u_{i_{2}}(\vec{x},\tau_2) u_{i_{3}}(\vec{x},\tau_3) } \meanBr{ u_i(\vec{x},t) u_{i_{1}}(\vec{x},\tau_1) } \frac{\dh^2 \meanBr{\lambda_{i_{1}}(\vec{x}, \tau_1) } }{\dh x_{i_{3}} \dh x_{i_{2}} } 
		\,.
	\end{split} \label{PS: eq: Fn series thrice second extra term general}
\end{align}

\subsection{Taylor expansion of the quasilinear term}
Expanding $\meanBr{\lambda_i}$ as indicated in equation \ref{PS.FOSA: taylor expansion Phi} allows us to evaluate the contributions arising from the first term on the RHS of equation \ref{PS: eq: series with terms up to v4}, neglecting $\bigO(\tau_c^2)$ terms:
\begin{align}
	\begin{split}
		\continuedTerm&\negphantom{\continuedTerm}
		- \velCorr{{\posTimeCombFree}_{i} {1}_{i_{1}}} \derG{i}{\posTimeCombFree|1} \meanBr{\lambda_{i_{1}}\posTimeComb{1} }
		\\
		={}&
		- \frac{\dh}{\dh x_i} \int_{-\infty}^t\d\tau \int\d \vec{q} \Dirac(\vec{x} - \vec{q}) \meanBr{ u_i(\vec{x},t) u_j(\vec{q},\tau) } \frac{\dh \Phi(\vec{q},t)}{\dh q_j}
		\\& - \frac{\dh}{\dh x_i} \int_{-\infty}^t\d\tau \int\d \vec{q} \Dirac(\vec{x} - \vec{q}) \meanBr{ u_i(\vec{x},t) u_j(\vec{q},\tau) } \left( \tau - t \right) \frac{\dh^2 \Phi(\vec{q},t)}{\dh q_j \dh t}
	\end{split}
	\\
	\begin{split}
		={}&
		- \frac{\dh}{\dh x_i} \int_{-\infty}^t\d\tau \meanBr{ u_i(\vec{x},t) u_j(\vec{x},\tau) } \frac{\dh \Phi(\vec{x},t)}{\dh x_j}
		\\& - \frac{\dh}{\dh x_i} \int_{-\infty}^t\d\tau \int_{-\infty}^t\d\tau_2 \meanBr{ u_i(\vec{x},t) u_j(\vec{x},\tau) } \left( \tau - t \right)
		\\&\continuedTerm\times \frac{\dh}{\dh x_j \dh x_k}{\left( \meanBr{ u_k(\vec{x},t) u_l(\vec{x},\tau_2) } \frac{\dh \Phi(\vec{x},t)}{\dh x_l} \right)}
	\end{split} \label{PS: eq: FOSA term Taylor O(tau) general}
\end{align}
where we have used equations \ref{PS: eq: Phi evolution} and \ref{PS: eq: turbulent diffusion term FOSA arbitrary velocity correlation}
(discarding $\kappa$ as before).

\subsection{Further simplification assuming separable correlations}
\label{PS: appendix: simplify separable correlator}

To simplify the expressions obtained above, we assume the velocity correlations are separable (equation \ref{PS: eq: separable correlator definition}).
From now on, we will discard terms which have more than two derivatives of the mean scalar (recall that $\meanBr{\lambda_i} = \dh_i\Phi$).
The temporal integrals over the correlation functions in the terms of interest can be written in terms of the constants $g_1$ and $g_2$, defined in equations \ref{PS: eq: g1 g2 defn}.
We write equation \ref{PS: eq: Fn series thrice first extra term general} as (omitting spatial and temporal arguments since they are the same for all terms)
\begin{align}
	\begin{split}
		\continuedTerm&\negphantom{\continuedTerm}
		- \velCorr{{\posTimeCombFree}_{i} {1}_{i_{1}}} \derG{i}{\posTimeCombFree|2} \velCorr{{2}_{i_{2}} {3}_{i_{3}}} \derG{i_{2}}{2|1} \derG{i_{1}}{1|3} \meanBr{\lambda_{i_{3}}\posTimeComb{3} }
		\\
		={}&
		- \tau_c g_1 \, \frac{\dh}{\dh x_i}{\left( C_{i i_1 i_2} C_{i_2 i_3 i_1} \meanBr{\lambda_{i_{3}} } \right)}
		- \tau_c g_1 \, \frac{\dh}{\dh x_i}{\left( C_{i i_1} C_{i_2, i_3, i_1, i_2} \meanBr{\lambda_{i_{3}} } \right)}
		\\& - \tau_c g_1 \, \frac{\dh}{\dh x_i}{\left( C_{i i_1} C_{i_2 i_3 i_1} \, \frac{\dh \meanBr{\lambda_{i_{3}} } }{\dh x_{i_{2}}} \right)}
		- \tau_c g_1 \, \frac{\dh}{\dh x_i}{\left( C_{i i_1 i_2} C_{i_2 i_3} \, \frac{\dh \meanBr{\lambda_{i_{3}} } }{\dh x_{i_{1}}} \right)}
		\\& - \tau_c g_1 \, \frac{\dh}{\dh x_i}{\left( C_{i i_1} C_{i_2 i_3 i_2} \, \frac{\dh \meanBr{\lambda_{i_{3}} } }{\dh x_{i_{1}}} \right)}
		\,.
	\end{split} \label{PS: eq: Fn series thrice first extra term intermediate 2}
\end{align}
Similarly, equation \ref{PS: eq: Fn series thrice second extra term general} becomes
\begin{align}
	\begin{split}
		\continuedTerm&\negphantom{\continuedTerm}
		- \velCorr{{\posTimeCombFree}_{i} {1}_{i_{1}}} \derG{i}{\posTimeCombFree|2} \velCorr{{2}_{i_{2}} {3}_{i_{3}}} \derG{i_{2}}{2|3} \derG{i_{3}}{3|1} \meanBr{\lambda_{i_{1}}\posTimeComb{1} }
		\\
		={}& - \tau_c g_2 \, \frac{\dh}{\dh x_i}{\left( C_{i i_1 i_3} C_{i_2 i_3 i_2} \meanBr{\lambda_{i_{1}} } \right)}
		- \tau_c g_2 \, \frac{\dh}{\dh x_i}{\left( C_{i_2 i_3} C_{i i_1 i_3 i_2} \meanBr{\lambda_{i_{1}} } \right)}
		\\& - \tau_c g_2 \, \frac{\dh}{\dh x_i}{\left( C_{i_2 i_3} C_{i i_1 i_3} \, \frac{\dh \meanBr{\lambda_{i_{1}} } }{\dh x_{i_{2}}} \right)}
		- \tau_c g_2 \, \frac{\dh}{\dh x_i}{\left( C_{i_2 i_3 i_2} C_{i i_1} \, \frac{\dh \meanBr{\lambda_{i_{1}} } }{\dh x_{i_{3}}} \right)}
		\\& - \tau_c g_2 \, \frac{\dh}{\dh x_i}{\left( C_{i_2 i_3} C_{i i_1 i_2} \, \frac{\dh \meanBr{\lambda_{i_{1}} } }{\dh x_{i_{3}}} \right)}
		\,.
	\end{split} \label{PS: eq: Fn series thrice second extra term intermediate 2}
\end{align}

We will now simplify the above using expressions from appendix \ref{PS: appendix: unequal-time-correlators separable} and keeping only up to one derivative of the turbulent quantities ($E$, $H$, and $N$), i.e.\@ we will neglect terms like $\nabla^2 E$ or $(\nabla E)^2$.
Equation \ref{PS: eq: Fn series thrice first extra term intermediate 2} becomes
\begin{align}
	\begin{split}
		\continuedTerm&\negphantom{\continuedTerm}
		- \velCorr{{\posTimeCombFree}_{i} {1}_{i_{1}}} \derG{i}{\posTimeCombFree|2} \velCorr{{2}_{i_{2}} {3}_{i_{3}}} \derG{i_{2}}{2|1} \derG{i_{1}}{1|3} \meanBr{\lambda_{i_{3}}\posTimeComb{3} }
		\\
		={}& - \frac{\tau_c g_1}{18} \, \frac{\dh}{\dh x_i} \left[ H^2 \, \frac{\dh\Phi}{\dh x_i} \right]
		\,.
	\end{split} \label{PS: eq: post-FOSA O(K) term 1}
\end{align}
Equation \ref{PS: eq: Fn series thrice second extra term intermediate 2} becomes
\begin{align}
	\begin{split}
		\continuedTerm&\negphantom{\continuedTerm}
		- \velCorr{{\posTimeCombFree}_{i} {1}_{i_{1}}} \derG{i}{\posTimeCombFree|2} \velCorr{{2}_{i_{2}} {3}_{i_{3}}} \derG{i_{2}}{2|3} \derG{i_{3}}{3|1} \meanBr{\lambda_{i_{1}}\posTimeComb{1} }
		\\
		={}& \frac{2 \tau_c g_2}{9} \frac{\dh}{\dh x_i} \left[ E N \frac{\dh\Phi}{\dh x_i} \right]
		\,.
	\end{split} \label{PS: eq: post-FOSA O(K) term 2}
\end{align}
Equation \ref{PS: eq: FOSA term Taylor O(tau) general} becomes
\begin{align}
	\begin{split}
		- \velCorr{{\posTimeCombFree}_{i} {1}_{i_{1}}} \derG{i}{\posTimeCombFree|1} \meanBr{\lambda_{i_{1}}\posTimeComb{1} }
		={}&
		- \frac{1}{3} \, \frac{\dh}{\dh x_i}{\left[ E \, \frac{\dh \Phi}{\dh x_i} \right]}
		\,.
	\end{split} \label{PS: eq: FOSA O(K) term}
\end{align}

One might get additional terms on keeping two spatial derivatives of the turbulent quantities, but for that, we would need expressions like those in appendix \ref{PS: appendix: unequal-time-correlators separable} up to the same order.

\section{\texorpdfstring{$n$}{n}-th functional derivative of \texorpdfstring{$B$}{B}}
\label{B: appendix: nth functional derivative of B}

Following a similar procedure to that used in appendix \ref{PS: appendix: nth functional derivative recursion-like relation}, we wish to derive an expression for the $n$-th functional derivative of $\vec{B}$ with respect to $\vec{u}$.
Denoting $\tilde{B}_i \defn B_i[u_m + \epsilon\chi_m]$ (where $\chi_m$ is some test function), we may use equation \ref{B: eq: B_i general equation in terms of GF} to write
\begin{equation}
	\tilde{B}_i(\vec{x},t) = \int\d\tau\d\vec{q}\, G^c(\vec{x},t|\vec{q},\tau) \, \epsilon_{ijk} \epsilon_{klm} \frac{\dh}{\dh q_j} \left[ \left( u_l(\vec{q},\tau) + \epsilon \chi_l(\vec{q},\tau)\right) \tilde{B}_m(\vec{q},\tau) \right]
	\,.
\end{equation}
Differentiating $n$ times and setting $\epsilon=0$, we obtain
\begin{align}
	\begin{split}
		\left. \frac{\d^n \tilde{B}_i(\vec{x},t) }{\d \epsilon^n} \right|_{\epsilon=0}
		={}&
		- \int\d\tau\d\vec{q}\, \epsilon_{ijk} \epsilon_{klm} \frac{\dh G^c(\vec{x},t|\vec{q},\tau) }{\dh q_j} \, u_l(\vec{q},\tau) \left. \frac{\d^{n} \tilde{B}_m(\vec{q},\tau) }{\d \epsilon^{n}} \right|_{\epsilon=0}
		\\& - n \int\d\tau\d\vec{q}\, \epsilon_{ijk} \epsilon_{klm} \frac{\dh G^c(\vec{x},t|\vec{q},\tau) }{\dh q_j} \, \chi_l(\vec{q},\tau) \left. \frac{\d^{n-1} \tilde{B}_m(\vec{q},\tau) }{\d \epsilon^{n-1}} \right|_{\epsilon=0}
		\,.
	\end{split}
\end{align}
Note that we have integrated by parts to shift the spatial derivative onto the Green function.
The functional derivatives of $B_i$ are then related by (where we use the notation $u_{j_i} \defn u_{j_{i}}(\vec{x}^{(i)},t^{(i)})$)
\begin{align}
	\begin{split}
		\frac{\delta^n B_i(\vec{x},t) }{\delta u_{i_1} \dots \delta u_{i_n} } 
		={}& - \int\d\tau\d\vec{q}\, \epsilon_{ijk} \epsilon_{klm} \frac{\dh G^c(\vec{x},t|\vec{q},\tau) }{\dh q_j} u_l(\vec{q},\tau) \frac{\delta^n B_m(\vec{q},\tau) }{\delta u_{i_1} \dots \delta u_{i_n} }
		\\& - \sum_{\alpha=1}^n  \epsilon_{ijk} \epsilon_{ki_{\alpha}m} \frac{\dh G^c(\vec{x},t|\vec{x}^{(\alpha)}, t^{(\alpha)}) }{\dh x^{(\alpha)}_j} \frac{\delta^{n-1} B_m(\vec{x}^{(\alpha)}, t^{(\alpha)}) }{\delta u_{i_1} \dots \delta u_{i_{\alpha-1}}\delta u_{i_{\alpha+1}} \dots \delta u_{i_n}} 
		\,.
	\end{split} \label{B: eq: n-th functional derivative of B without averaging}
\end{align}
Averaging both sides of the above and applying the Furutsu-Novikov theorem, we obtain
\begin{align}
	\begin{split}
		\meanBr{ \frac{\delta^n B_i(\vec{x},t) }{\delta u_{i_1} \dots \delta u_{i_n} } }
		={}& - \int\d\tau\d\vec{q}\d\vec{x}^{(n+1)}\d t^{(n+1)}\, \bigg[ \epsilon_{ijk} \epsilon_{klm} \frac{\dh G^c(\vec{x},t|\vec{q},\tau) }{\dh q_j}
		\\&\continuedTerm\times  \meanBr{ u_l(\vec{q},\tau) u_{i_{n+1}}(\vec{x}^{(n+1)}, t^{(n+1)}) } \meanBr{ \frac{\delta^{n+1} B_m(\vec{q},\tau) }{\delta u_{i_1} \dots \delta u_{i_n} \delta u_{i_{n+1}} } }  \bigg]
		\\& - \sum_{\alpha=1}^n \epsilon_{ijk} \epsilon_{ki_{\alpha}m} \frac{\dh G^c(\vec{x},t|\vec{x}^{(\alpha)}, t^{(\alpha)}) }{\dh x^{(\alpha)}_j} \meanBr{ \frac{\delta^{n-1} B_m(\vec{x}^{(\alpha)}, t^{(\alpha)}) }{\delta u_{i_1} \dots \delta u_{i_{\alpha-1}}\delta u_{i_{\alpha+1}} \dots \delta u_{i_n}} }
		\,.
	\end{split} \label{B: eq: n-th functional derivative of B recursion-like relation}
\end{align}

\section{Simplification of the EMF retaining \texorpdfstring{$\bigO(\tau_c)$}{O(tau)} terms}
\label{B: appendix: simplify FuruNovi O(tau) terms}
For simplicity, we assume $\eta=0$, in which case the diffusion Green function becomes a positional Dirac delta.\footnote{
Note that $\int_{-\infty}^\infty f(x)g(y) \frac{\dh}{\dh x} \delta(x-y) \,\d x$ should be evaluated as $- f'(y) g(y)$, and not $- \frac{\dh}{\dh y} \left[ f(y) g(y) \right]$.
}

\subsection{Quasilinear term}
We write (neglecting $\bigO(\tau_c^2)$ terms)\footnote{
We are not Taylor-expanding the Green function since we plan to set $\eta=0$, where we have $G(\vec{x},t|\vec{q},\tau) = \Dirac(\vec{x} - \vec{q}) \Heaviside(t - \tau)$.
}
\begin{align}
	\begin{split}
		\velCorr{{\posTimeCombFree}_{i} {1}_{i_{1}}} \derG{i_{2}}{\posTimeCombFree|1} \Bbar_{i_{3}}\posTimeComb{1}
		={}&
		\velCorr{{\posTimeCombFree}_{i} {1}_{i_{1}}} \derG{i_{2}}{\posTimeCombFree|1} \Bbar_{i_{3}}(\vec{x}^{(1)}, t)
		\\& + \velCorr{{\posTimeCombFree}_{i} {1}_{i_{1}}} \derG{i_{2}}{\posTimeCombFree|1} \left( t^{(1)} - t \right) \frac{\dh \Bbar_{i_{3}}(\vec{x}^{(1)}, t) }{\dh t}
		\,.
	\end{split}
\end{align}
We recall that neglecting $\bigO(\tau_c)$ terms and setting $\eta = 0$, one can write (equations \ref{B.FN: eq: white noise EMF} and \ref{B.FN: eq: mean field induction general EMF})
\begin{equation}
	\frac{\dh B_a}{\dh t}
	=
	\epsilon_{abc} \left( - \frac{H}{6} \dh_b \Bbar_{c} - \frac{E}{3} \epsilon_{cde} \dh_b \dh_d \Bbar_e \right)
	\,.
\end{equation}
Using this and dropping terms with more than one spatial derivative of $\Bbar$, we write
\begin{align}
	\begin{split}
		\continuedTerm&\negphantom{\continuedTerm}
		\velCorr{{\posTimeCombFree}_{i} {1}_{i_{1}}} \derG{i_{2}}{\posTimeCombFree|1} \Bbar_{i_{3}}\posTimeComb{1}
		\\
		={}&
		\velCorr{{\posTimeCombFree}_{i} {1}_{i_{1}}} \derG{i_{2}}{\posTimeCombFree|1} \Bbar_{i_{3}}(\vec{x}^{(1)}, t)
		\\& - \velCorr{{\posTimeCombFree}_{i} {1}_{i_{1}}} \derG{i_{2}}{\posTimeCombFree|1} \left( t^{(1)} - t \right) \epsilon_{i_3 bc} \, \frac{H}{6} \, \frac{\dh \Bbar_{c}{(\vec{x}^{(1)}, t)} }{\dh x^{(1)}_b}
	\end{split}
	\\
	\begin{split}
		={}&
		- \int_{ t^{(1)} } \meanBr{ u_i(\vec{x}, t) \, \frac{\dh u_{i_1}(\vec{x}, t^{(1)}) }{\dh x_{i_2} } } \Heaviside(t - t^{(1)}) \, \Bbar_{i_{3}}(\vec{x}, t)
		\\& - \int_{ t^{(1)} } \meanBr{ u_i(\vec{x}, t) \, u_{i_1}(\vec{x}, t^{(1)}) } \Heaviside(t - t^{(1)}) \, \frac{\dh \Bbar_{i_{3}}(\vec{x}, t) }{\dh x_{i_2} }
		\\& - \frac{H}{6} \, \epsilon_{i_3 bc} \int_{ t^{(1)} } \meanBr{ u_i(\vec{x}, t) \, \frac{\dh u_{i_1}(\vec{x}, t^{(1)}) }{\dh x_{i_2} } } \Heaviside(t - t^{(1)}) \left( t^{(1)} - t \right) \frac{\dh \Bbar_{c}{(\vec{x}, t)} }{\dh x_b}
		\,.
	\end{split}
\end{align}
Assuming the velocity correlation is separable (equation \ref{PS: eq: separable correlator definition}), we write
\begin{align}
	\begin{split}
		\velCorr{{\posTimeCombFree}_{i} {1}_{i_{1}}} \derG{i_{2}}{\posTimeCombFree|1} \Bbar_{i_{3}}\posTimeComb{1}
		={}&
		- \frac{1}{2} C_{i i_1 i_2} \Bbar_{i_{3}}
		- \frac{1}{2} C_{i i_1} \frac{\dh \Bbar_{i_{3}} }{\dh x_{i_2} }
		+ \frac{\tau_c H}{12} \, \epsilon_{i_3 bc} C_{i i_1 i_2} \frac{\dh \Bbar_{c} }{\dh x_b}
		\,.
	\end{split}
\end{align}
Using the results of appendix \ref{PS: appendix: unequal-time-correlators separable}, we then write
the contribution to the EMF ($\emf_k = \epsilon_{kij} \left< V_i B_j \right>$) as
\begin{align}
	\begin{split}
		- \epsilon_{kij} \Upsilon_{j i_2 i_1 i_3} \velCorr{{\posTimeCombFree}_{i} {1}_{i_{1}}} \derG{i_{2}}{\posTimeCombFree|1} \Bbar_{i_{3}}\posTimeComb{1}
		={}&
		- \frac{H}{6} \, B{}_{k}
		+ \left(\frac{E}{3} - \frac{H^{2} \tau_{c}}{36}\right)\epsilon{}_{kr_{0}r_{1}} B{}_{r_{0}, r_{1}}
	\end{split} \label{B: eq: FOSA O(tau) term emf contribution simplified}
\end{align}
where we have used a comma to denote differentiation.

\subsection{First higher-order term}

Denoting $\funnyInt \dots \defn \int \d\tau_1\d\tau_2\tau_3 \, \Heaviside(t-\tau_2)  \Heaviside(\tau_2-\tau_3)  \Heaviside(\tau_3-\tau_1) $;
dropping terms involving more than one spatial derivative of $\Bbar$;\footnote{
The term that appears in the induction equation is $\curl \vec{\emf}$.
}
and assuming we are only interested in homogeneous turbulence, we write
\begin{align}
	\begin{split}
		\continuedTerm&\negphantom{\continuedTerm}
		\velCorr{{\posTimeCombFree}_{i} {1}_{i_{1}}} \derG{j_{3}}{\posTimeCombFree|2} \velCorr{{2}_{i_{2}} {3}_{i_{3}}} \derG{i_{4}}{2|3} \derG{i_{6}}{3|1} \Bbar_{i_{7}}\posTimeComb{1} 
		\\
		={}& 
		- \funnyInt u_i^{(1)}(t) u_{i_2}^{(2)}(\tau_2) \frac{\dh u_{i_3}^{(2)}(\tau_3) }{\dh x_{i_4} } \frac{\dh }{\dh x_{j_3} } \bigg[ \frac{\dh u_{i_1}^{(1)}(\tau_1) }{\dh x_{i_6} } \Bbar_{i_{7}}(\tau_1) \bigg]
		\\&
		- \funnyInt u_i^{(1)}(t) u_{i_2}^{(2)}(\tau_2) u_{i_3}^{(2)}(\tau_3) \frac{\dh }{\dh x_{j_3} } \bigg[ \frac{\dh^2 u_{i_1}^{(1)}(\tau_1) }{ \dh x_{i_6} \dh x_{i_4} } \Bbar_{i_{7}}(\tau_1) \bigg]
		\\&
		- \funnyInt u_i^{(1)}(t) u_{i_2}^{(2)}(\tau_2) u_{i_3}^{(2)}(\tau_3)  \frac{\dh^2 u_{i_1}^{(1)}(\tau_1) }{\dh x_{i_6} \dh x_{j_3} } \frac{\dh \Bbar_{i_{7}}(\tau_1) }{\dh x_{i_4} }
		\\&
		- \funnyInt u_i^{(1)}(t) u_{i_2}^{(2)}(\tau_2) \frac{\dh u_{i_3}^{(2)}(\tau_3) }{\dh x_{i_4} } \frac{\dh u_{i_1}^{(1)}(\tau_1) }{\dh x_{j_3} } \frac{\dh \Bbar_{i_{7}}(\tau_1) }{\dh x_{i_6} } 
		\\&
		- \funnyInt u_i^{(1)}(t) u_{i_2}^{(2)}(\tau_2) u_{i_3}^{(2)}(\tau_3) \frac{\dh^2 u_{i_1}^{(1)}(\tau_1) }{\dh x_{i_4} \dh x_{j_3} } \frac{\dh \Bbar_{i_{7}}(\tau_1) }{\dh x_{i_6} } 
	\end{split}
\end{align}
where, since the position argument in all the terms of the final expression is $\vec{x}$, we have not indicated it explicitly.
Above, as long as one is willing to ignore $\bigO(\tau_c^2)$ terms, one can replace the time arguments of all occurrences of $\Bbar$ by $t$.
Assuming the correlation of the velocity field is separable (equation \ref{PS: eq: separable correlator definition}), we can write
\begin{align}
	\begin{split}
		\continuedTerm&\negphantom{\continuedTerm}
		\velCorr{{\posTimeCombFree}_{i} {1}_{i_{1}}} \derG{j_{3}}{\posTimeCombFree|2} \velCorr{{2}_{i_{2}} {3}_{i_{3}}} \derG{i_{4}}{2|3} \derG{i_{6}}{3|1} \Bbar_{i_{7}}\posTimeComb{1}
		\\
		={}& 
		- \tau_c g_2 C_{i i_1 j_3 i_6} C_{i_2 i_3 i_4} \Bbar_{i_{7}}
		- \tau_c g_2 C_{i i_1 i_6} C_{i_2 i_3 i_4} \frac{\dh \Bbar_{i_{7}} }{\dh x_{j_3} }
		\\&
		- \tau_c g_2 C_{i i_1 j_3 i_6 i_4} C_{i_2 i_3} \Bbar_{i_{7}}
		- \tau_c g_2 C_{i i_1 i_6 i_4} C_{i_2 i_3} \frac{\dh \Bbar_{i_{7}} }{\dh x_{j_3} }
		\\&
		- \tau_c g_2 C_{i i_1 i_6 j_3} C_{i_2 i_3} \frac{\dh \Bbar_{i_{7}} }{\dh x_{i_4} }
		- \tau_c g_2 C_{i i_1 j_3} C_{i_2 i_3 i_4} \frac{\dh \Bbar_{i_{7}} }{\dh x_{i_6} }
		\\&
		- \tau_c g_2 C_{i i_1 i_4 j_3} C_{i_2 i_3} \frac{\dh \Bbar_{i_{7}} }{\dh x_{i_6} }
	\end{split}
\end{align}
where $g_2$ is defined in equation \ref{PS: eq: g1 g2 defn}.
Using expressions for the tensors $C_{ij\dots}$ from appendix \ref{PS: appendix: unequal-time-correlators separable}, we write the contribution to the EMF ($\emf_k = \epsilon_{kij} \left< V_i B_j \right>$) as
\begin{align}
	\begin{split}
		\continuedTerm&\negphantom{\continuedTerm}
		- \epsilon_{kij} \Upsilon_{i_{5} i_{6} i_{1} i_{7}} \Upsilon_{j j_{3} i_{2} j_{2}} \Upsilon_{j_{2} i_{4} i_{3} i_{5}} \velCorr{{\posTimeCombFree}_{i} {1}_{i_{1}}} \derG{j_{3}}{\posTimeCombFree|2} \velCorr{{2}_{i_{2}} {3}_{i_{3}}} \derG{i_{4}}{2|3} \derG{i_{6}}{3|1} \Bbar_{i_{7}}\posTimeComb{1}
		\\
		={}&
		-\left(\frac{2 \tau_c E N g_{2}}{9} + \frac{\tau_c H^{2} g_{2}}{9}\right)\epsilon{}_{kr_{0}r_{1}} B{}_{r_{0}, r_{1}} + \left(\frac{2 \tau_c E L g_{2}}{9} + \frac{\tau_c H N g_{2}}{9}\right) B{}_{k}
		\,.
	\end{split} \label{B: eq: Furunovi O(tau) term 1 simplified}
\end{align}

\subsection{Second higher-order term}

Denoting $\funnyInt \dots \defn \int \d\tau_1\d\tau_2\tau_3 \, \Heaviside(t-\tau_2)  \Heaviside(\tau_2-\tau_1)  \Heaviside(\tau_1-\tau_3) \dots $;
dropping terms with more than one spatial derivative of $\Bbar$;
and assuming we are interested in homogeneous turbulence, we write
\begin{align}
	\begin{split}
		\continuedTerm&\negphantom{\continuedTerm}
		\velCorr{{\posTimeCombFree}_{i} {1}_{i_{1}}} \derG{j_{3}}{\posTimeCombFree|2} \velCorr{{2}_{i_{2}} {3}_{i_{3}}} \derG{i_{4}}{2|1} \derG{i_{6}}{1|3} \Bbar_{i_{7}}\posTimeComb{3} 
		\\
		={}& - \funnyInt u_{i_2}^{(2)}(\tau_2) \frac{\dh u_{i_3}^{(2)}(\tau_3) }{\dh x_{i_6} } u_i^{(1)}(t) \frac{\dh  }{\dh x_{j_3} } \left[  \frac{\dh u_{i_1}^{(1)}(\tau_1) }{\dh x_{i_4} } \Bbar_{i_{7}}(\tau_3) \right]
		\\& 
		- \funnyInt u_{i_2}^{(2)}(\tau_2) \frac{\dh^2 u_{i_3}^{(2)}(\tau_3) }{\dh x_{i_6} \dh x_{i_4} } u_i^{(1)}(t) \frac{\dh  }{\dh x_{j_3} } \left[  u_{i_1}^{(1)}(\tau_1)  \Bbar_{i_{7}}(\tau_3) \right]
		\\& 
		- \funnyInt u_{i_2}^{(2)}(\tau_2) \frac{\dh u_{i_3}^{(2)}(\tau_3) }{\dh x_{i_6} } u_i^{(1)}(t) \frac{\dh u_{i_1}^{(1)}(\tau_1) }{\dh x_{j_3} } \frac{\dh \Bbar_{i_{7}}(\tau_3) }{\dh x_{i_4} }
		\\&
		- \funnyInt u_i^{(1)}(t) u_{i_2}^{(2)}(\tau_2) u_{i_3}^{(2)}(\tau_3) \frac{\dh^2 u_{i_1}^{(1)}(\tau_1) }{\dh x_{i_4} \dh x_{j_3} } \frac{\dh  \Bbar_{i_{7}}(\tau_3) }{\dh x_{i_6} }
		\\&
		- \funnyInt u_{i_2}^{(2)}(\tau_2) \frac{\dh u_{i_3}^{(2)}(\tau_3) }{\dh x_{i_4} } u_i^{(1)}(t) \frac{\dh u_{i_1}^{(1)}(\tau_1) }{\dh x_{j_3} } \frac{\dh  \Bbar_{i_{7}}(\tau_3) }{\dh x_{i_6} }
		\,.
	\end{split}
\end{align}
Above, as long as one is willing to ignore $\bigO(\tau_c^2)$ terms, one can replace the time arguments of all occurrences of $\Bbar$ by $t$.
Assuming the correlation of the velocity field is separable (equation \ref{PS: eq: separable correlator definition}), we can write
\begin{align}
	\begin{split}
		\continuedTerm&\negphantom{\continuedTerm}
		\velCorr{{\posTimeCombFree}_{i} {1}_{i_{1}}} \derG{j_{3}}{\posTimeCombFree|2} \velCorr{{2}_{i_{2}} {3}_{i_{3}}} \derG{i_{4}}{2|1} \derG{i_{6}}{1|3} \Bbar_{i_{7}}\posTimeComb{3}
		\\
		={}&
		- \tau_c g_1 C_{i_2 i_3 i_6} C_{i i_1 i_4 j_3} \Bbar_{i_{7}}
		- \tau_c g_1 C_{i_2 i_3 i_6} C_{i i_1 i_4} \frac{\dh \Bbar_{i_{7}} }{\dh x_{j_3} }
		\\&
		- \tau_c g_1 C_{i_2 i_3 i_6 i_4} C_{i i_1 j_3} \Bbar_{i_{7}}
		- \tau_c g_1 C_{i_2 i_3 i_6 i_4} C_{i i_1} \frac{\dh \Bbar_{i_{7}} }{\dh x_{j_3} }
		- \tau_c g_1 C_{i_2 i_3 i_6} C_{i i_1 j_3} \frac{\dh \Bbar_{i_{7}} }{\dh x_{i_4} }
		\\&
		- \tau_c g_1 C_{i i_1 i_4 j_3} C_{i_2 i_3} \frac{\dh  \Bbar_{i_{7}} }{\dh x_{i_6} }
		- \tau_c g_1 C_{i_2 i_3 i_4} C_{i i_1 j_3} \frac{\dh  \Bbar_{i_{7}} }{\dh x_{i_6} }
	\end{split}
\end{align}
where $g_1$ is defined in equation \ref{PS: eq: g1 g2 defn}.
Using expressions for the tensors $C_{ij\dots}$ from appendix \ref{PS: appendix: unequal-time-correlators separable}, we write the contribution to the EMF ($\emf_k = \epsilon_{kij} \left< V_i B_j \right>$) as
\begin{align}
	\begin{split}
		\continuedTerm&\negphantom{\continuedTerm}
		- \epsilon_{kij} \Upsilon_{i_{5} i_{6} i_{3} i_{7}} \Upsilon_{j j_{3} i_{2} j_{2}} \Upsilon_{j_{2} i_{4} i_{1} i_{5}} 
		\velCorr{{\posTimeCombFree}_{i} {1}_{i_{1}}} \derG{j_{3}}{\posTimeCombFree|2} \velCorr{{2}_{i_{2}} {3}_{i_{3}}} \derG{i_{4}}{2|1} \derG{i_{6}}{1|3} \Bbar_{i_{7}}\posTimeComb{3}
		\\
		={}&
		-\left(\frac{\tau_c H^2 g_1}{18}\right) \epsilon{}_{kr_{0}r_{1}}B{}_{r_{0},r_{1}}
		\,.
	\end{split} \label{B: eq: Furunovi O(tau) term 2 simplified}
\end{align}

\end{document}